\documentclass[aps, prc, twocolumn,preprintnumbers,amsmath,amssymb,longbibliography,superscriptaddress]{revtex4-2}

\usepackage[T1]{fontenc}
\usepackage{graphicx}
\usepackage{dcolumn}
\usepackage{bm}
\usepackage{url}
\usepackage{hyperref}
\begin{document}


\title{Border of the Island of Inversion: Unbound states in $^{29}$Ne}

\author {M.~Holl}
\author {S.~Lindberg}
\author {A.~Heinz}
\affiliation{Institutionen f\"{o}r Fysik, Chalmers Tekniska H\"{o}gskola, 412 96 G\"{o}teborg, Sweden}
\author {Y.~Kondo}
\author {T.~Nakamura}
\affiliation{Department of Physics, Tokyo Institute of Technology, 2-12-1 O-Okayama, Meguro, Tokyo, 152-8551, Japan}
\author {J.~A.~Tostevin}
\affiliation{Department of Physics, Tokyo Institute of Technology, 2-12-1 O-Okayama, Meguro, Tokyo, 152-8551, Japan}
\affiliation{Department of Physics, Faculty of Engineering and Physical Sciences, University of Surrey, Guildford, Surrey GU2 7XH , United Kingdom}
\author {H.~Wang}
\affiliation{Department of Physics, Tokyo Institute of Technology, 2-12-1 O-Okayama, Meguro, Tokyo, 152-8551, Japan}
\author {T.~Nilsson}
\affiliation{Institutionen f\"{o}r Fysik, Chalmers Tekniska H\"{o}gskola, 412 96 G\"{o}teborg, Sweden}
\author {N.~L.~Achouri}
\affiliation{LPC-Caen, ENSICAEN, CNRS/IN2P3, UNICAEN, Normandie Universit\'e, 14050 Caen, France}
\author {H.~Al Falou}
\affiliation{Lebanese University, Beirut, Lebanon}
\author {L.~Atar}
\affiliation{Institut f\"ur Kernphysik, Technische Universit\"at Darmstadt, 64289 Darmstadt, Germany}
\author {T.~Aumann}
\affiliation{Institut f\"ur Kernphysik, Technische Universit\"at Darmstadt, 64289 Darmstadt, Germany}
\affiliation{GSI Helmholtzzentrum f\"ur Schwerionenforschung, 64291 Darmstadt, Germany}
\author{H.~Baba}
\affiliation{RIKEN Nishina Center, Hirosawa 2-1, Wako, Saitama 351-0198, Japan}
\author{K.~Boretzky}
\affiliation{GSI Helmholtzzentrum f\"ur Schwerionenforschung, 64291 Darmstadt, Germany}
\author{C.~Caesar}
\affiliation{Institut f\"ur Kernphysik, Technische Universit\"at Darmstadt, 64289 Darmstadt, Germany}
\affiliation{GSI Helmholtzzentrum f\"ur Schwerionenforschung, 64291 Darmstadt, Germany}
\author{D.~Calvet}
\affiliation{Irfu, CEA, Universit\'e Paris-Saclay, 91191 Gif-sur-Yvette, France} 
\author{H.~Chae}
\affiliation{Department of Physics and Astronomy, Seoul National University, 1 Gwanak-ro, Gwanak-gu, Seoul 08826, Republic of Korea} 
\author{N.~Chiga}
\affiliation{RIKEN Nishina Center, Hirosawa 2-1, Wako, Saitama 351-0198, Japan}
\author{A.~Corsi}
\affiliation{Irfu, CEA, Universit\'e Paris-Saclay, 91191 Gif-sur-Yvette, France} 
\author{H.~L.~Crawford}
\affiliation{Nuclear Science Division, Lawrence Berkeley National Laboratory, Berkeley, California 94720, USA}
\author{F.~Delaunay}
\affiliation{LPC-Caen, ENSICAEN, CNRS/IN2P3, UNICAEN, Normandie Universit\'e, 14050 Caen, France}
\author{A.~Delbart}
\affiliation{Irfu, CEA, Universit\'e Paris-Saclay, 91191 Gif-sur-Yvette, France} 
\author{Q.~Deshayes}
\affiliation{LPC-Caen, ENSICAEN, CNRS/IN2P3, UNICAEN, Normandie Universit\'e, 14050 Caen, France}
\author{P.~D\'iaz~Fern\'andez}
\affiliation{Institutionen f\"{o}r Fysik, Chalmers Tekniska H\"{o}gskola, 412 96 G\"{o}teborg, Sweden}
\author{Z.~Dombr\'adi}
\affiliation{Institute of Nuclear Research, Atomki, 4001 Debrecen, Hungary}
\author{C.~A.~Douma}
\affiliation{ESRIG, University of Groningen, Zernikelaan 25, 9747 AA Groningen, The Netherlands} 
\author{Z.~Elekes}
\affiliation{Institute of Nuclear Research, Atomki, 4001 Debrecen, Hungary}
\author{P.~Fallon}
\affiliation{Nuclear Science Division, Lawrence Berkeley National Laboratory, Berkeley, California 94720, USA}
\author{I.~Ga\v{s}pari\'{c}}
\affiliation{Ru\dj er Bo\v skovi\'c Institute, HR-10000 Zagreb, Croatia}
\affiliation{RIKEN Nishina Center, Hirosawa 2-1, Wako, Saitama 351-0198, Japan}
\affiliation{Institut f\"ur Kernphysik, Technische Universit\"at Darmstadt, 64289 Darmstadt, Germany}
\author{J.-M.~Gheller}
\affiliation{Irfu, CEA, Universit\'e Paris-Saclay, 91191 Gif-sur-Yvette, France} 
\author{J.~Gibelin}
\affiliation{LPC-Caen, ENSICAEN, CNRS/IN2P3, UNICAEN, Normandie Universit\'e, 14050 Caen, France}
\author{A.~Gillibert}
\affiliation{Irfu, CEA, Universit\'e Paris-Saclay, 91191 Gif-sur-Yvette, France} 
\author{M.~N.~Harakeh}
\affiliation{GSI Helmholtzzentrum f\"ur Schwerionenforschung, 64291 Darmstadt, Germany}
\affiliation{ESRIG, University of Groningen, Zernikelaan 25, 9747 AA Groningen, The Netherlands} 
\author{A.~Hirayama}
\affiliation{Department of Physics, Tokyo Institute of Technology, 2-12-1 O-Okayama, Meguro, Tokyo, 152-8551, Japan}
\author{C.~R.~Hoffman}
\affiliation{Physics Division, Argonne National Laboratory, Argonne, Illinois 60439, USA}
\author{A.~Horvat}
\affiliation{GSI Helmholtzzentrum f\"ur Schwerionenforschung, 64291 Darmstadt, Germany}
\author{\'A.~Horv\'ath}
\affiliation{E\"otv\"os Lor\'and University, P\'azm\'any P\'eter S\'et\'any 1/A, H-1117 Budapest, Hungary}
\author{J.~W.~Hwang}
\affiliation{Center for Exotic Nuclear Studies, Institute for Basic Science, Daejeon 34126, Republic of Korea}
\affiliation{Department of Physics and Astronomy, Seoul National University, 1 Gwanak-ro, Gwanak-gu, Seoul 08826, Republic of Korea} 
\author{T.~Isobe}
\affiliation{RIKEN Nishina Center, Hirosawa 2-1, Wako, Saitama 351-0198, Japan}
\author{J.~Kahlbow}
\affiliation{Institut f\"ur Kernphysik, Technische Universit\"at Darmstadt, 64289 Darmstadt, Germany}
\affiliation{RIKEN Nishina Center, Hirosawa 2-1, Wako, Saitama 351-0198, Japan}
\author{N.~Kalantar-Nayestanaki}
\affiliation{ESRIG, University of Groningen, Zernikelaan 25, 9747 AA Groningen, The Netherlands} 
\author{S.~Kawase}
\affiliation{Department of Advanced Energy Engineering Science, Kyushu University, Kasuga, Fukuoka 816-8580, Japan}
\author{S.~Kim}
\affiliation{Center for Exotic Nuclear Studies, Institute for Basic Science, Daejeon 34126, Republic of Korea}
\affiliation{Department of Physics and Astronomy, Seoul National University, 1 Gwanak-ro, Gwanak-gu, Seoul 08826, Republic of Korea} 
\author{K.~Kisamori}
\affiliation{RIKEN Nishina Center, Hirosawa 2-1, Wako, Saitama 351-0198, Japan}
\author{T.~Kobayashi}
\affiliation{Department of Physics, Tohoku University, Miyagi 980-8578, Japan} 
\author{D.~K\"{o}rper}
\affiliation{GSI Helmholtzzentrum f\"ur Schwerionenforschung, 64291 Darmstadt, Germany}
\author{S.~Koyama}
\affiliation{University of Tokyo, Tokyo 1130033, Japan}
\author{I.~Kuti}
\affiliation{Institute of Nuclear Research, Atomki, 4001 Debrecen, Hungary}
\author{V.~Lapoux}
\affiliation{Irfu, CEA, Universit\'e Paris-Saclay, 91191 Gif-sur-Yvette, France} 
\author{F.~M.~Marqu\'es}
\affiliation{LPC-Caen, ENSICAEN, CNRS/IN2P3, UNICAEN, Normandie Universit\'e, 14050 Caen, France}
\author{S.~Masuoka}
\affiliation{Center for Nuclear Study, University of Tokyo, 2-1 Hirosawa, Wako, Saitama 351-0198, Japan}
\author{J.~Mayer}
\affiliation{Institut für Kernphysik, Universität zu Köln, 50937 Köln, Germany}
\author{K.~Miki}
\affiliation{National Superconducting Cyclotron Laboratory, Michigan State University, East Lansing, Michigan 48824, USA}
\author{T.~Murakami}
\affiliation{Department of Physics, Kyoto University, Kyoto 606-8502, Japan}
\author{M.~Najafi}
\affiliation{ESRIG, University of Groningen, Zernikelaan 25, 9747 AA Groningen, The Netherlands} 
\author{K.~Nakano}
\affiliation{Department of Advanced Energy Engineering Science, Kyushu University, Kasuga, Fukuoka 816-8580, Japan}
\author{N.~Nakatsuka}
\affiliation{Department of Physics, Kyoto University, Kyoto 606-8502, Japan}
\author{A.~Obertelli}
\affiliation{Irfu, CEA, Universit\'e Paris-Saclay, 91191 Gif-sur-Yvette, France} 
\author{F.~de Oliveira Santos}
\affiliation{Grand Acc\'el\'erateur National d’Ions Lourds (GANIL), CEA/DRF-CNRS/IN2P3, Bvd Henri Becquerel, 14076 Caen, France}
\author{N.~A.~Orr}
\affiliation{LPC-Caen, ENSICAEN, CNRS/IN2P3, UNICAEN, Normandie Universit\'e, 14050 Caen, France}
\author{H.~Otsu}
\affiliation{RIKEN Nishina Center, Hirosawa 2-1, Wako, Saitama 351-0198, Japan}
\author{T.~Ozaki}
\affiliation{Department of Physics, Tokyo Institute of Technology, 2-12-1 O-Okayama, Meguro, Tokyo, 152-8551, Japan}
\author{V.~Panin}
\affiliation{RIKEN Nishina Center, Hirosawa 2-1, Wako, Saitama 351-0198, Japan}
\author{S.~Paschalis}
\affiliation{Institut f\"ur Kernphysik, Technische Universit\"at Darmstadt, 64289 Darmstadt, Germany}
\author{A.~Revel}
\affiliation{Grand Acc\'el\'erateur National d’Ions Lourds (GANIL), CEA/DRF-CNRS/IN2P3, Bvd Henri Becquerel, 14076 Caen, France}
\author{D.~Rossi}
\affiliation{Institut f\"ur Kernphysik, Technische Universit\"at Darmstadt, 64289 Darmstadt, Germany}
\author{A.~T.~Saito}
\affiliation{Department of Physics, Tokyo Institute of Technology, 2-12-1 O-Okayama, Meguro, Tokyo, 152-8551, Japan}
\author{T.~Y.~Saito}
\affiliation{University of Tokyo, Tokyo 1130033, Japan}
\author{M.~Sasano}
\author{H.~Sato}
\affiliation{RIKEN Nishina Center, Hirosawa 2-1, Wako, Saitama 351-0198, Japan}
\author{Y.~Satou}
\affiliation{Department of Physics and Astronomy, Seoul National University, 1 Gwanak-ro, Gwanak-gu, Seoul 08826, Republic of Korea} 
\author{H.~Scheit}
\author{F.~Schindler}
\affiliation{Institut f\"ur Kernphysik, Technische Universit\"at Darmstadt, 64289 Darmstadt, Germany}
\author{P.~Schrock}
\affiliation{Center for Nuclear Study, University of Tokyo, 2-1 Hirosawa, Wako, Saitama 351-0198, Japan}
\author{M.~Shikata}
\affiliation{Department of Physics, Tokyo Institute of Technology, 2-12-1 O-Okayama, Meguro, Tokyo, 152-8551, Japan}
\author{Y.~Shimizu}
\affiliation{RIKEN Nishina Center, Hirosawa 2-1, Wako, Saitama 351-0198, Japan}
\author{H.~Simon}
\affiliation{GSI Helmholtzzentrum f\"ur Schwerionenforschung, 64291 Darmstadt, Germany}
\author{D.~Sohler}
\affiliation{Institute of Nuclear Research, Atomki, 4001 Debrecen, Hungary}
\author{O.~Sorlin}
\affiliation{Grand Acc\'el\'erateur National d’Ions Lourds (GANIL), CEA/DRF-CNRS/IN2P3, Bvd Henri Becquerel, 14076 Caen, France}
\author{L.~Stuhl}
\affiliation{RIKEN Nishina Center, Hirosawa 2-1, Wako, Saitama 351-0198, Japan}
\affiliation{Center for Exotic Nuclear Studies, Institute for Basic Science, Daejeon 34126, Republic of Korea}
\author{S.~Takeuchi}
\affiliation{Department of Physics, Tokyo Institute of Technology, 2-12-1 O-Okayama, Meguro, Tokyo, 152-8551, Japan}
\author{M.~Tanaka}
\affiliation{Department of Physics, Osaka University, Osaka 560-0043, Japan}
\author{M.~Thoennessen}
\affiliation{National Superconducting Cyclotron Laboratory, Michigan State University, East Lansing, Michigan 48824, USA}
\author{H.~T\"{o}rnqvist}
\affiliation{Institut f\"ur Kernphysik, Technische Universit\"at Darmstadt, 64289 Darmstadt, Germany}
\author{Y.~Togano}
\affiliation{Department of Physics, Tokyo Institute of Technology, 2-12-1 O-Okayama, Meguro, Tokyo, 152-8551, Japan}
\affiliation{Department of Physics, Rikkyo University, Tokyo 171-8501, Japan}
\author{T.~Tomai}
\affiliation{Department of Physics, Tokyo Institute of Technology, 2-12-1 O-Okayama, Meguro, Tokyo, 152-8551, Japan}
\author{J.~Tscheuschner}
\affiliation{Institut f\"ur Kernphysik, Technische Universit\"at Darmstadt, 64289 Darmstadt, Germany}
\author{J.~Tsubota}
\affiliation{Department of Physics, Tokyo Institute of Technology, 2-12-1 O-Okayama, Meguro, Tokyo, 152-8551, Japan}
\author{T.~Uesaka}
\author{Z.~Yang}
\affiliation{RIKEN Nishina Center, Hirosawa 2-1, Wako, Saitama 351-0198, Japan}
\author{M.~Yasuda}
\affiliation{Department of Physics, Tokyo Institute of Technology, 2-12-1 O-Okayama, Meguro, Tokyo, 152-8551, Japan}
\author{K.~Yoneda}
\affiliation{RIKEN Nishina Center, Hirosawa 2-1, Wako, Saitama 351-0198, Japan}

\collaboration{SAMURAI21 collaboration}\noaffiliation

\date{\today}


  
\begin{abstract}
The nucleus $^{29}$Ne is situated at the border of the island of inversion. Despite significant efforts, no bound low-lying intruder $f_{7/2}$-state, which would place $^{29}$Ne firmly inside the island of inversion, has yet been observed. Here, the first investigation of unbound states of $^{29}$Ne is reported. The states were populated in $^{30}\mathrm{Ne}(p,pn)$ and  $^{30}\mathrm{Na}(p,2p)$ reactions at a beam energy of around $230$~MeV/nucleon, and analyzed in terms of their resonance properties, partial cross sections and momentum distributions. The momentum distributions are compared to calculations using the eikonal, direct reaction model, allowing $\ell$-assignments for the observed states. The lowest-lying resonance at an excitation energy of 1.48(4)~MeV shows clear signs of a significant $\ell$=3-component, giving first evidence for $f_{7/2}$ single particle strength in $^{29}$Ne. The excitation energies and strengths of the observed states are compared to shell-model calculations using the {\sc sdpf-u-mix} interaction.
\end{abstract}

\maketitle
\section{Introduction}
Understanding the evolution of shell structure when moving away from stability and towards the driplines is a key challenge for modern nuclear physics. In neutron-rich nuclei, the mismatch between neutron and proton numbers leads to a quenching of shell gaps and the onset of deformation. In particular, in neutron-rich Ne, Na and Mg nuclei, the collapse of the $N=20$ and $N=28$ shell gaps can be observed. The ground states of several isotopes of these nuclei are found to be dominated by intruder configurations from the $1f_{7/2}$ and $2p_{3/2}$ orbitals. In terms of the nuclear shell model, this can be explained by the influence of the tensor force of the nucleon-nucleon interaction. The resulting reduced attraction between the proton $\pi d_{5/2}$ and neutron $\nu d_{3/2}$ orbitals leads to the latter rising in energy, closing the shell gaps at and above $N$=20 \cite{Otsuka2020}.\\
This so-called island-of-inversion has received a lot of experimental attention and considerable efforts have been made to map the exact boundaries of this region. Initially believed to be restricted to $20\leq N \leq 22$ \cite{Warburton1990,Orr1991}, the transition to the island of inversion has since been shown to happen gradually in a large number of experiments, see e.g.\ Refs.~\cite{Obertelli2006, Terry2006, Catford2010, Matta2019}. In the case of the neon isotopic chain, an intruder state was observed in $^{27}$Ne at low excitation energy \cite{Obertelli2006,Terry2006,Brown2012}. That this state was also observed in $1n$-knockout from $^{28}$Ne indicates that $p_{3/2}$ intruder configurations also contribute to the ground state of $^{28}$Ne \cite{Terry2006}. Isotopes with larger $N$, were firmly placed inside the island of inversion. For the even-$N$ isotopes $^{30,32}$Ne, low-lying first $2^+$-states and a large enhancement of collectivity were observed \cite{Doornenbal2009,Doornenbal2016,Murray2019}. In addition, $1n$-removal reactions from $^{29,31}$Ne revealed large $p_{3/2}$ contributions to the ground state configurations of these isotopes \cite{Kobayashi2016,Liu2017,Nakamura2009,Nakamura2014}. The isotope $^{29}$Ne is therefore lying at the border of the island of inversion. However, the gradual increase of the influence of intruder states along the isotopic chain warrants more detailed studies into the evolution of single-particle orbitals.\\
While information on the excitation energies of the lowest lying $p_{3/2}$ states exists for neutron-rich Ne isotopes across the island of inversion, this is not the case for the $f_{7/2}$ states. Low-lying $f_{7/2}$-states provide another clear sign of the collapse of the $N$=20 shell gap. However, while such states have been observed for some isotopes, e.g. $^{27}$Ne \cite{Brown2012}, they have not yet been observed systematically. The location of the $f_{7/2}$-state relative to the $p_{3/2}$-state is also important in evaluating the $N$=28 gap and may provide a clue for understanding its collapse in neutron-rich Mg and Si isotopes.\\
In the case of $^{29}$Ne, no evidence for bound $f_{7/2}$ strength was found in a recent study of one-neutron removal from $^{30}$Ne \cite{Liu2017}, leading to the assumption that this strength leads to states above the neutron decay threshold.\\
In the present work, we report a first investigation of unbound states of $^{29}$Ne. These states were populated using two different single-nucleon removal reactions, $^{30}\mathrm{Ne}(p,pn)$ and  $^{30}\mathrm{Na}(p,2p)$, studied in kinematically complete measurements, and analyzed in terms of their resonance properties, partial cross sections and momentum distributions. The combination of these results gives a detailed insight into the structure of $^{29}$Ne. While  $^{30}\mathrm{Ne}(p,pn)$ allows investigation of the influence of $pf$-intruder states at the border of the island of inversion, $^{30}\mathrm{Na}(p,2p)$ gives complimentary information on states populated through removal of a deeply-bound proton. The experimental data also provide a test of state-of-the-art shell model calculations.
\section{Experiment}
\begin{figure}[t]
\includegraphics[width=0.49\textwidth]{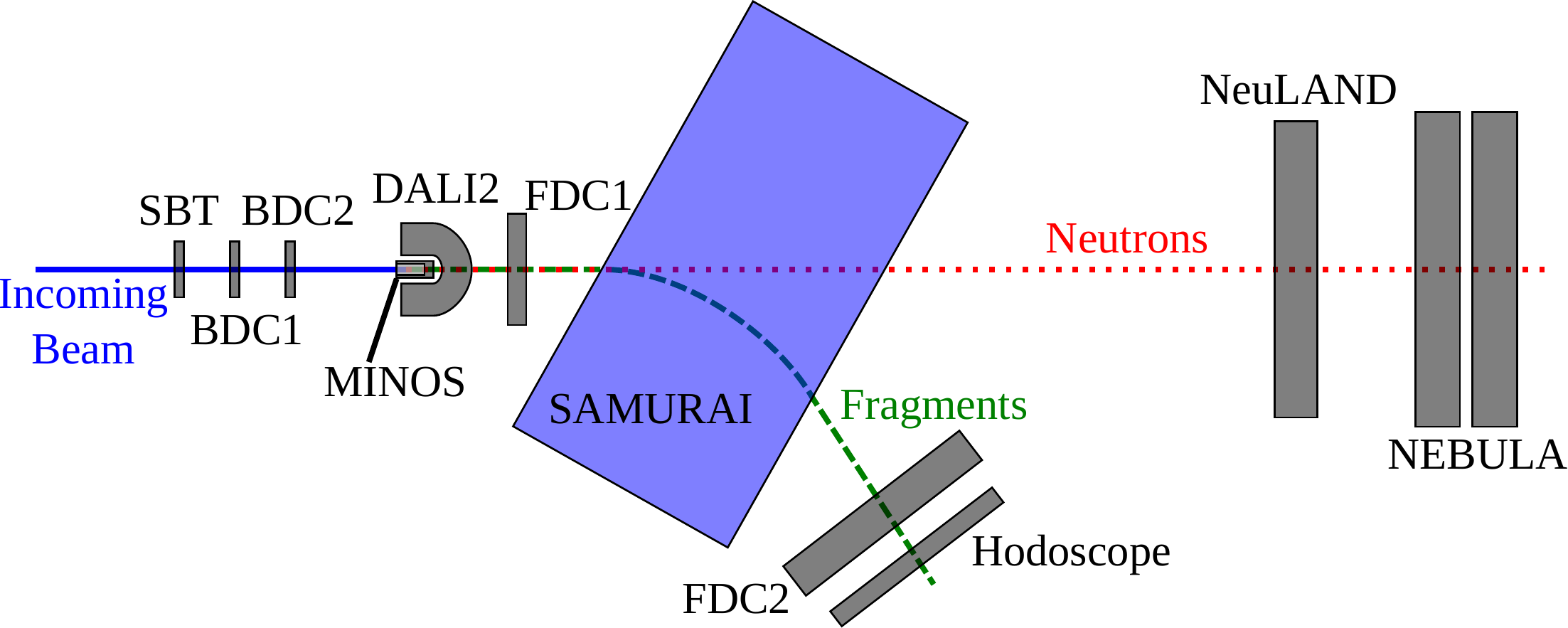}
\caption{Schematic overview of the SAMURAI setup. The setup is able to detect and identify the incoming beam particles, outgoing fragments and scattered protons from $(p,2p)$ and $(p,pn)$ reactions, as well as $\gamma$-rays and neutrons emitted in-flight by de-exciting fragments. See text for details. }
\label{fig:setup}
\end{figure}
The experiment was carried out at the Radioactive Isotope Beam Factory (RIBF) operated by the RIKEN Nishina Center and the Center for Nuclear Study (CNS), University of Tokyo. Radioactive beams of $^{30}$Ne and $^{30}$Na were produced by fragmentation of a 345 MeV/nucleon $^{48}$Ca beam impinging on a 2.8~g/cm$^2$ beryllium target with a typical intensity of 500 pnA. They were then selected and transported to the SAMURAI setup using the BigRIPS fragment separator \cite{Kubo2003}. To purify the beams, achromatic aluminum wedge degraders were inserted at the focal planes F1 and F5 of BigRIPS, with thicknesses of 15~mm and 7~mm, respectively. The data presented here was taken concurrently with a study of $^{28}$F, presented in Ref. \cite{Revel2020}. Therefore, two different settings of the separator were used, one centered on $^{29}$Ne and one on $^{29}$F. The $^{30}$Ne beam had a mid-target energy of 240~MeV/nucleon and an intensity of 250 particles per second. For the $^{30}$Na beam, the energy and intensity were 229 MeV/nucleon and 120 particles per second.\\
A schematic overview of the SAMURAI setup is shown in Fig. \ref{fig:setup}. The incoming ions were identified by their charge and time-of-flight using a set of two 0.5~mm thin plastic scintillators (SBT) while two multi-wire drift chambers (BDC1 and BDC2) were used to track their path towards the MINOS liquid hydrogen target \cite{Obertelli2014}. The target cell of MINOS is 151 mm long and its entrance window has a diameter of 38 mm. The target cell was cooled to 15~K and the pressure inside the cell was 143 mbar, leading to a target thickness of 1.1~g/cm$^2$. The cell is surrounded by a time-projection-chamber used for measuring the outgoing protons from $(p,2p)$ and $(p,pn)$ reactions. Combining the information of the tracks of the incoming beam with those of the outgoing protons allows to reconstruct the reaction vertex with a precision of $\sigma=3$ mm. The time-projection-chamber itself was surrounded by the $\gamma$-ray detector array DALI2 \cite{Takeuchi2014} consisting of 142 NaI(Tl) crystals. DALI2 was used to detect $\gamma$-rays emitted by de-exciting fragments in flight. For $E_{\gamma}=1$~MeV, the photo-peak efficiency of DALI2 was $\epsilon_{\gamma}\approx15\%$ and the resolution after Doppler correction was $\sigma\approx50$~keV. The charged reaction fragments were deflected by the magnetic field of the SAMURAI superconducting dipole magnet \cite{Kobasyashi2013} with 2.9~T in the center. The dipole gap of SAMURAI was kept under vacuum using a chamber with thin exit windows \cite{Shimizu2013}. Their positions before and after the magnet were measured using multi-wire drift chambers (FDC1 and FDC2). Together with their charge and time-of-flight, measured by a hodoscope consisting of 24 plastic scintillator bars at the end of the beamline, this allowed to fully reconstruct the four-vectors of the fragments.\\
Neutrons emitted in-flight were detected using the combination of the NeuLAND \cite{Boretzky2021} demonstrator and NEBULA \cite{Nakamura2016,Kondo2020}. The NeuLAND demonstrator consisted of 400 plastic scintillator bars arranged in 4 double planes, with each double plane consisting of a horizontal and a vertical plane. The 120 scintillator paddles of NEBULA are split into two walls, with each wall consisting of two vertical layers. The two large area detector arrays were placed at distances of 11 m and 14 m from the target, respectively.
\section{Analysis}
\begin{figure}[t]
\includegraphics[width=0.49\textwidth]{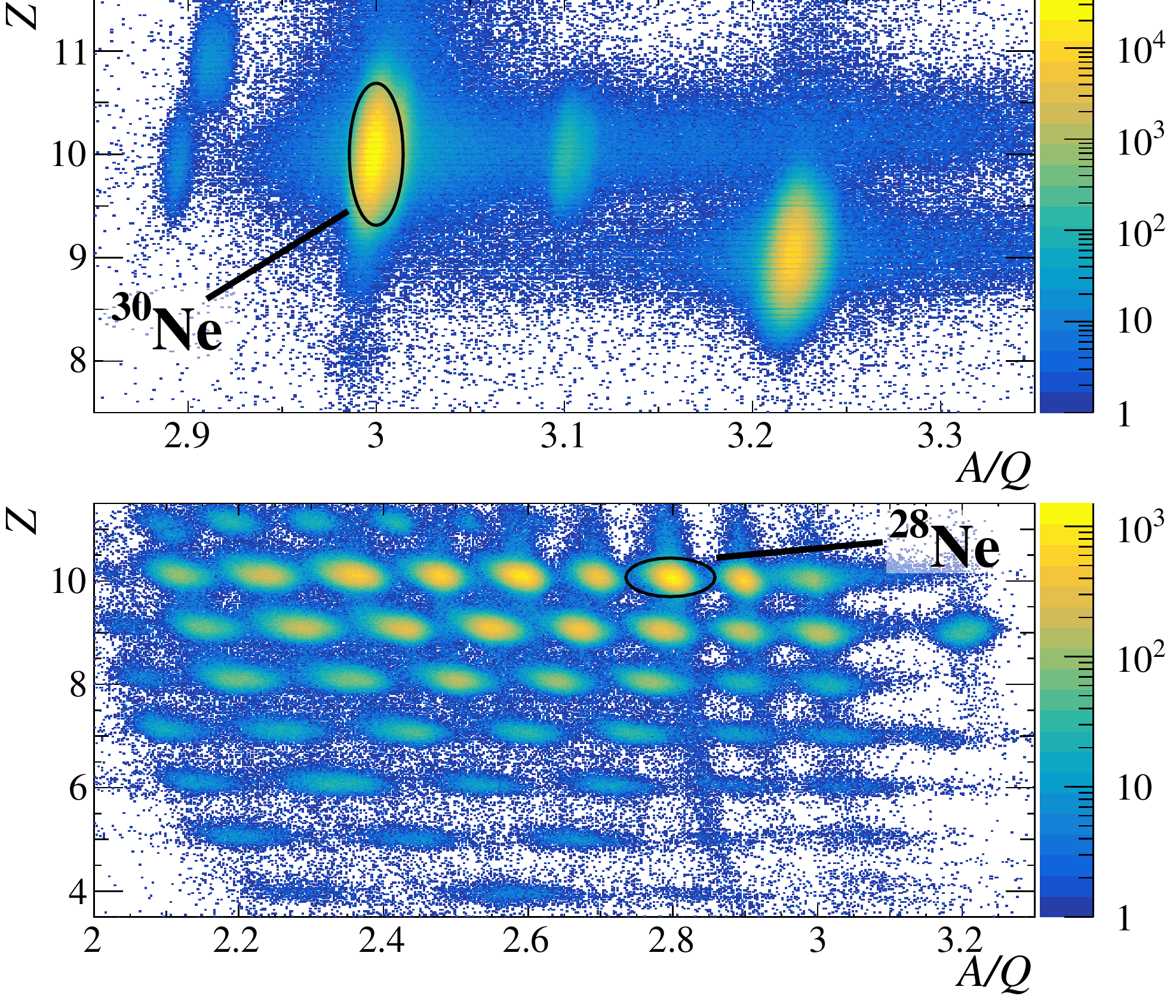}
\caption{Particle identification for $^{30}\mathrm{Ne}(p,pn)$. Upper panel: Incoming particle identification using the charge $Z$ and mass-to-charge-ratio ($A/Q$) measured by BigRIPS and the SBT. Lower panel: Outgoing particle identification using the charge measured by the hodoscope and the $A/Z$ extracted from the time-of-flight and flight path through the magnetic field of SAMURAI. Events with incoming $^{30}$Ne and one reconstructed track in MINOS have been selected. The $^{28}$Ne fragments created by neutron emission from excited $^{29}$Ne are marked in the figure.}
\label{fig:pid}
\end{figure}
\begin{figure}[t]
\includegraphics[width=0.49\textwidth]{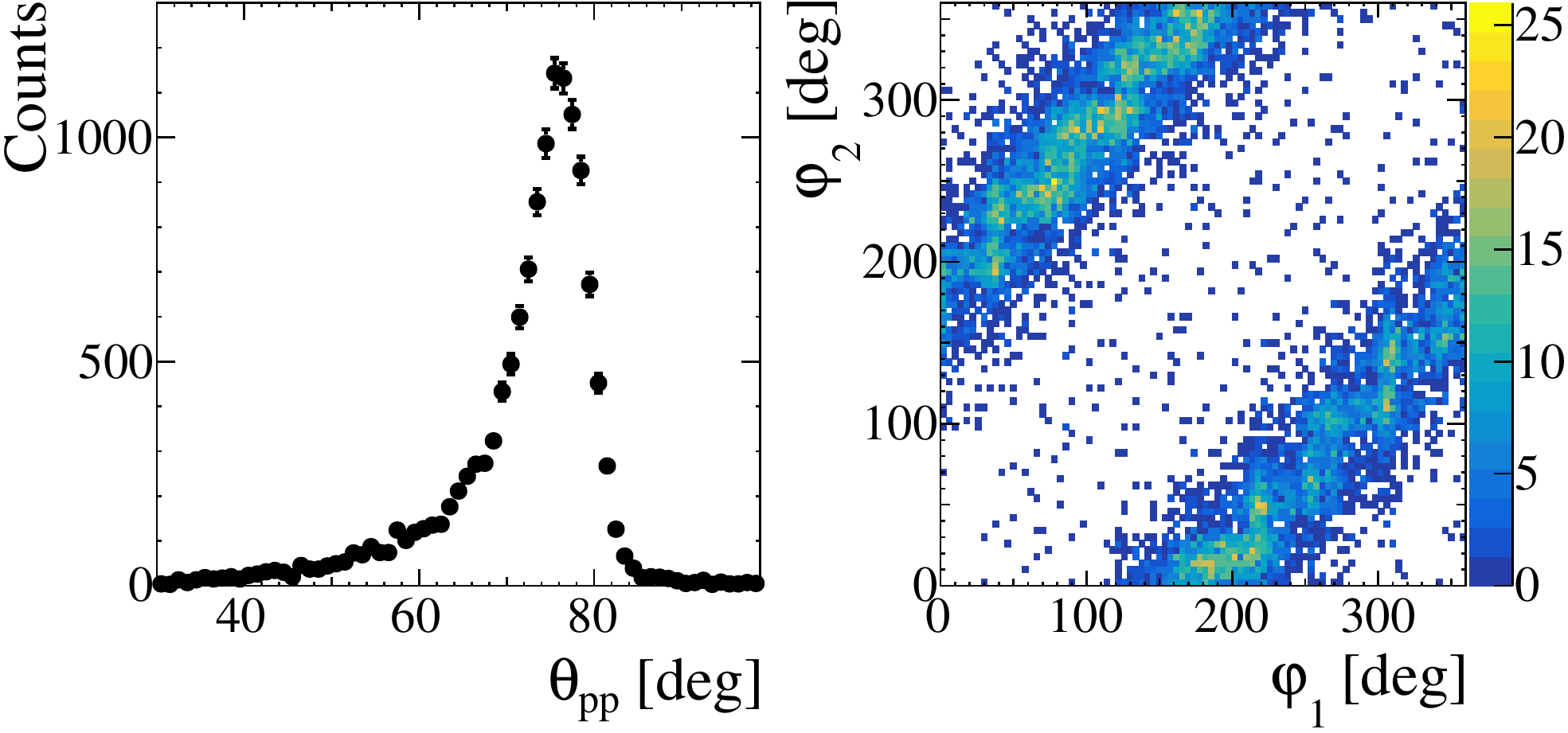}
\caption{Left: Opening angle $\theta_{pp}$ between the two protons detected by the MINOS TPC following $^{30}\mathrm{Na}(p,2p)$. A peak is observed slightly below $80^{\circ}$, characteristic for quasi-free scattering of a deeply-bound proton. Right: The azimuthal angles $\phi_{1,2}$ of the two protons plotted against each other. The two bands demonstrate the coplanar reaction kinematics.}
\label{fig:opang}
\end{figure}
The $^{30}\mathrm{Ne}(p,pn)$ and $^{30}\mathrm{Na}(p,2p)$ reactions were analyzed in the same manner. The incoming isotopes were identified using their charge $Z$ and mass-to-charge-ratio ($A/Q$) as extracted from the $B\rho$, time-of-flight and $\Delta E$ measurements by BigRIPS standard detectors and the SBT at the entrance of the setup. The upper panel of Fig. \ref{fig:pid} shows $A/Q$ and $Z$ plotted against each other for the case of $^{30}$Ne. The incoming beam was selected by fitting a two-dimensional Gaussian distribution to this plot and selecting particles within a $3\sigma$-interval. The same technique was applied to the reaction fragments, using the charge measured by the hodoscope while the $A/Z$ was determined from their $B\rho$, time-of-flight and flight-path through the magnetic field of SAMURAI. The identification plot for the outgoing fragments is shown in the lower panel of Fig. \ref{fig:pid} with incoming $^{30}$Ne and one particle track in MINOS selected. Again particles within a $3\sigma$ interval were chosen. To identify events from $(p,2p)$ and $(p,pn)$ reactions, at least one reconstructed track in the MINOS TPC was required.\\
While this is not a fully exclusive measurement of quasi-free scattering, the recoil protons are partially detected in $^{30}\mathrm{Na}(p,2p)$ and the recoil neutron is not detected in  $^{30}\mathrm{Ne}(p,pn)$, it is reasonable to assume that the quasi-free scattering mechanism dominates. To demonstrate this, the opening angle between the two protons, $\theta_{pp}$, and their $\phi$-angle correlations are shown in Fig. \ref{fig:opang} for events where two protons were detected in MINOS following $^{30}\mathrm{Na}(p,2p)$. The opening angle $\theta_{pp}$ shows a clear peak slightly below $80^{\circ}$, characteristic of quasi-free scattering of a deeply-bound proton. The two bands visible in the plot of the $\phi$-angles demonstrate the dominance of coplanar reaction kinematics, also characteristic of quasi-free scattering. The ratio between events with one and two protons detected is in line with what is expected from the efficiency of MINOS \cite{Obertelli2014} and indicates that the majority of events with just one reconstructed proton track also originate from quasi-free scattering. It can be assumed that the same holds true in the case of $^{30}\mathrm{Ne}(p,pn)$ where less-bound neutrons are removed.\\
The four-momenta of neutrons emitted from the $^{29}$Ne reaction fragments were determined using their hit positions in NeuLAND or NEBULA and the time-of-flight between these detectors and the reconstructed reaction vertex in the target. The four-momenta of both neutron $(E_n,\vec{p}_n)$ and $^{28}$Ne fragment $(E_f,\vec{p}_f)$ were used to reconstruct the invariant mass $M_{inv}$ of the unbound $^{29}$Ne system,
\begin{equation}
    \label{eq:minv}
    M_{inv} = \sqrt{(E_f+E_n)^2-\left|\vec{p}_f+\vec{p}_n\right|^2}. 
\end{equation}
From the invariant mass, the relative energy $E_{rel}$ can be extracted as 
\begin{equation}
    \label{eq:erel}
    E_{rel} = M_{inv}-m_f-m_n, 
\end{equation}
where the neutron and $^{28}$Ne masses are represented by $m_n$ and $m_f$, respectively.\\
The efficiency and resolution of the  $E_{rel}$ reconstruction was determined with Monte Carlo simulations. Using the experimental distributions for beam energy and target position as input, decay events of $^{29}$Ne were generated with varying relative energies in the range of $0-10$ MeV. The detector response to these events was then obtained using GEANT4 simulations including the entire experimental geometry and interactions of a neutron in NeuLAND and NEBULA. The combined acceptance and efficiency obtained this way is shown in the inset of the middle panel of Fig. \ref{fig:erel}. 
The energy resolution (full width at half maximum) FWHM for $E_{rel}$ was found to vary according to
\begin{equation}
    \label{eq:fwhm}
    \mathrm{FWHM}(E_{rel}) = (0.16\cdot E_{rel}^{0.74} + 0.06)~\mathrm{MeV}. 
\end{equation} 
The GEANT4 simulations were also used to obtain the detector response of DALI2 to $\gamma$-ray transitions of different energies.\\
To calculate the reaction cross sections, the number of incoming particles was approximated by the number of unreacted $^{30}$Ne or $^{30}$Na nuclei detected by the fragment arm of the SAMURAI setup, thus making corrections for the efficiencies of the fragment detectors and losses after the target unnecessary. Losses due to (secondary) reactions in the target were taken into account for both unreacted and reacted beam by GEANT4 simulations using the INCL/ABLA model \cite{Boudard2013}.
\section{Results}
\subsection{Relative Energy Spectra}
\begin{figure}[t]
\includegraphics[width=0.49\textwidth]{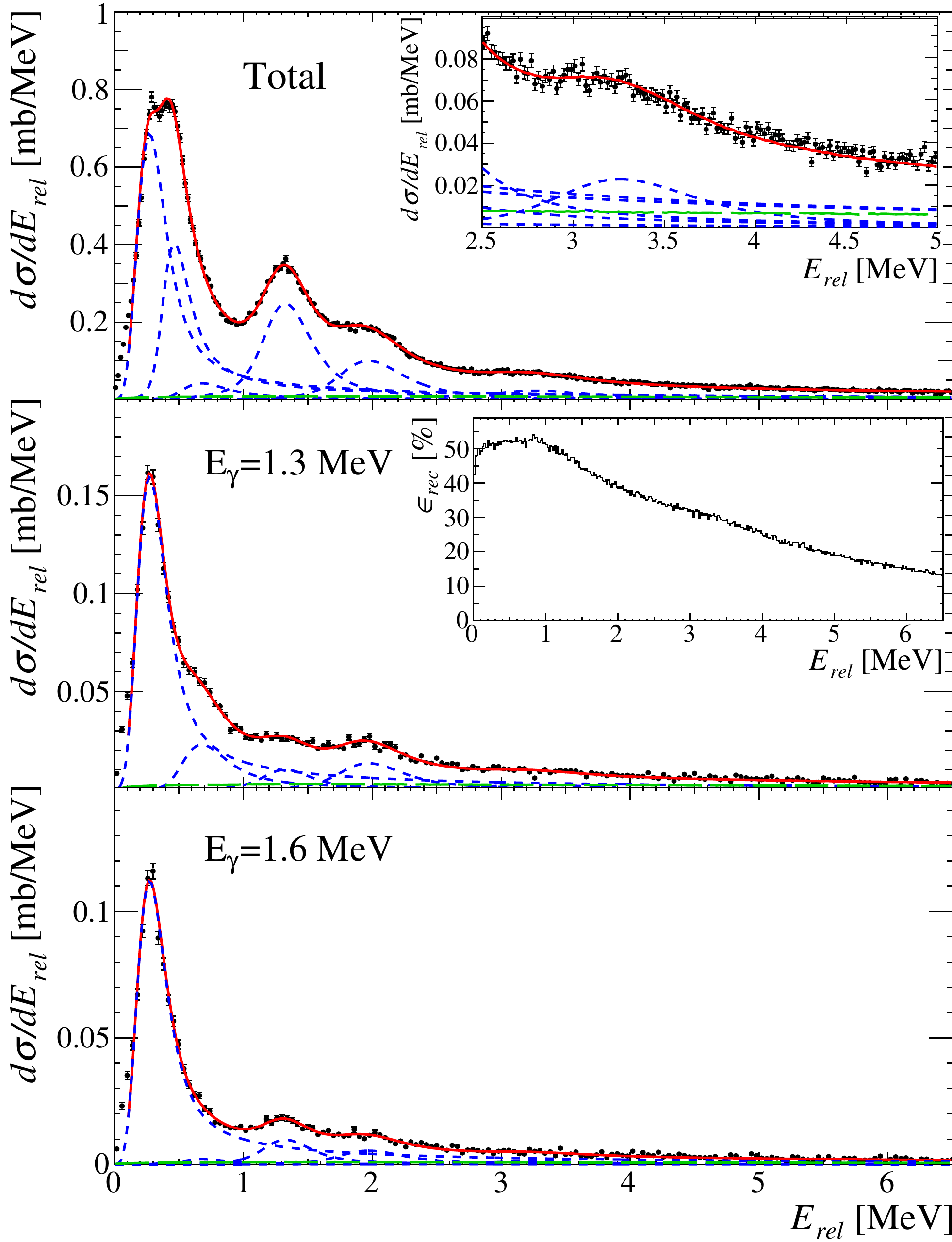}
\caption{Relative energy spectra of $^{28}$Ne$+n$ following $^{30}$Ne($p,pn$). The total fit (red line) consists of a combination of six Breit-Wigner resonances (blue dashed lines) and a non-resonant background (green dotted line). Upper panel: Total $E_{rel}$ spectrum. The inset shows a zoom on the energy region 2.5~MeV$<E_{rel}<$5~MeV. Middle panel: $E_{rel}$ spectrum gated on the $E_{\gamma}=1.3$ MeV transition. The inset shows the reconstruction efficiency $\epsilon_{rec}$. Lower panel: $E_{rel}$ spectrum gated on the $E_{\gamma}=1.6$ MeV transition.}
\label{fig:erel}
\end{figure}
\begin{figure}[t]
\includegraphics[width=0.49\textwidth]{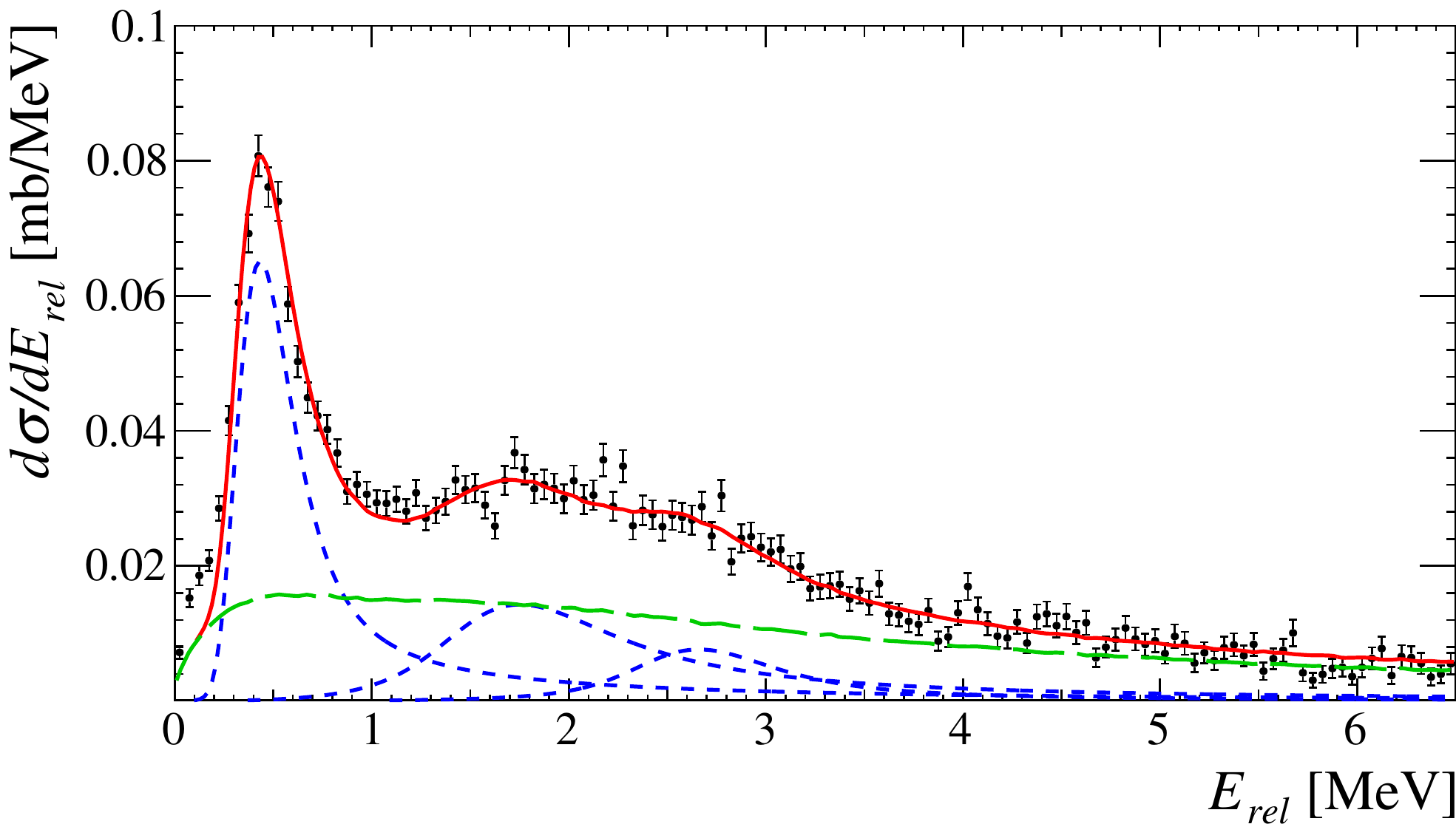}
\caption{Relative energy spectra of $^{28}$Ne$+n$ following $^{30}$Na($p,2p$). The total fit (red line) consists of a combination of three Breit-Wigner resonances (blue dashed lines) and a non-resonant background (green dotted line).}
\label{fig:erel_p2p}
\end{figure}
The relative energy spectrum of the unbound $^{29}$Ne system following $^{30}\mathrm{Ne}(p,pn)$ is shown in Fig. \ref{fig:erel}. The top panel shows the total $E_{rel}$ spectrum, whereas in the middle and bottom panel the spectrum is presented with gates on the photo peaks of the $\gamma$-ray transitions with $E_{\gamma}=1.3$~MeV and $E_{\gamma}=1.6$~MeV, respectively (see Sec. \ref{sec:egamma} for details).\\
To determine the properties of the observed peaks, the spectrum was fitted using a combination of energy-dependent Breit-Wigner line shapes of the form 
\begin{equation}
    \frac{d\sigma}{dE_{rel}} \propto \frac{\Gamma}{(E_{rel}-(E_{r}+\Delta_{\ell}(E_{rel})))^2+\Gamma^2/4},
\end{equation}
with the resonance energy $E_r$, the width of the resonance 
\begin{equation}
    \Gamma = \Gamma_r\frac{P_{\ell}(E_{rel})}{P_{\ell}(E_{r})}, 
\end{equation}
and the shift
\begin{equation}
    \Delta_{\ell}= \Gamma_r\frac{S_{\ell}(E_{r})-S_{\ell}(E_{rel})}{2\cdot P_{\ell}(E_{r})}, 
\end{equation}
with the intrinsic width $\Gamma_r$, the penetrability $P_{\ell}$ and the shift function $S_{\ell}$ \cite{Lane1958}. To include the experimental resolution, the resonance shapes were folded with a Gaussian with varying width according to Eq. \ref{eq:fwhm}. In addition, a non-resonant background obtained from event-mixing was included in the fit \cite{Marques2000, Randisi2014}. As can be seen in Fig. 4, this background is found to be almost negligible in the case of $^{30}\mathrm{Ne}(p,pn)$. Six resonances were identified and their positions and widths were determined. The $\ell$-assignments necessary to calculate $P_{\ell}$ were made using the analysis of the longitudinal and transverse momentum distributions, see Sec. \ref{sec:momentum}. Comparing the spectra with and without $\gamma$-ray coincidences, the large peak around $E_{rel}$=0.5~MeV is shown to consist of three resonances at 0.32(2)~MeV, 0.51(4)~MeV and 0.74(7)~MeV with intrinsic widths $\Gamma_r$ of 0.05(1)~MeV($\ell$=2), 0.17(8)~MeV($\ell$=2) and 0.11(5)~MeV($\ell$=2), respectively. Three more resonances are observed at 1.35(2)~MeV ($\Gamma_r=0.30(4)$~MeV, $\ell$=1), 2.01(1)~MeV($\Gamma_r=0.33(12)$~MeV, $\ell$=2) and 3.31(8)~MeV($\Gamma_r=0.7(3)$~MeV, $\ell$=2).\\
The $E_{rel}$-spectrum following $^{30}\mathrm{Na}(p,2p)$ is shown in Fig.~\ref{fig:erel_p2p}. The spectrum can be described with a non-resonant background and 3 resonances with $\ell$=2 at 0.48(4)~MeV, 1.95(18)~MeV and 2.72(12)~MeV. The intrinsic widths $\Gamma_r$ are found to be 0.05(3)~MeV, 0.8(4)~MeV and 0.6(4)~MeV, respectively.\\
For both reactions, the energy region $E_{rel}<0.15$~MeV was excluded from the fit. This was done because at these low $E_{rel}$, the detector response becomes asymmetric and can no longer be described as a Gaussian. Resonances at these energies can therefore also no longer be described as a Breit-Wigner folded with a Gaussian. Neither the analysis from coincident $\gamma$-rays, nor the momentum distributions of the fragments give any indication of additional resonances at these low $E_{rel}$. Varying the threshold $E_{rel}$ from which to include events changes the cross sections of the two resonances with the lowest $E_{rel}$ by $\sim$2~\%. This effect is included in the stated uncertainty of these values.
\subsection{Gamma-Ray Energy Spectra} \label{sec:egamma}
\begin{figure}[t]
\includegraphics[width=0.49\textwidth]{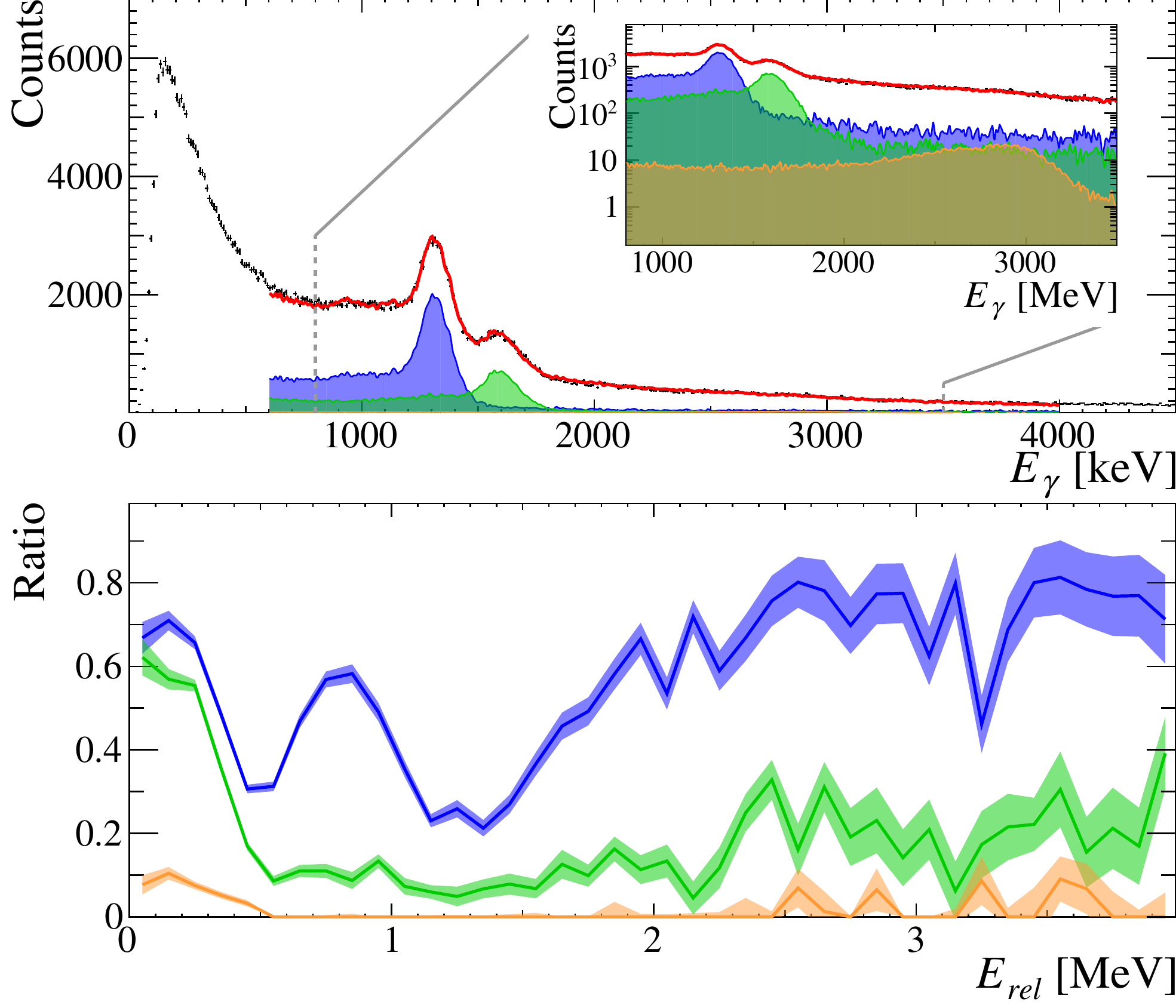}
\caption{Upper panel: Total $E_{\gamma}$-spectrum measured with DALI2. The spectrum is fitted by a combination of an exponential background and simulated line shapes for eight transitions. The contributions from the three transitions with $E_{\gamma}=1.3$~MeV (blue), $E_{\gamma}=1.6$~MeV (green) and $E_{\gamma}=2.9$~MeV (yellow) are highlighted. The inset shows a section of the spectrum in logarithmic scale. Lower panel:  Ratio of the efficiency-corrected integral of each simulated line shape and the total number of $^{28}$Ne$+n$ events for 50~keV-wide $E_{rel}$ bins for the three transitions highlighted above. The line width corresponds to the uncertainty.}
\label{fig:gamma}
\end{figure}
\begin{figure}[t]
\includegraphics[width=0.49\textwidth]{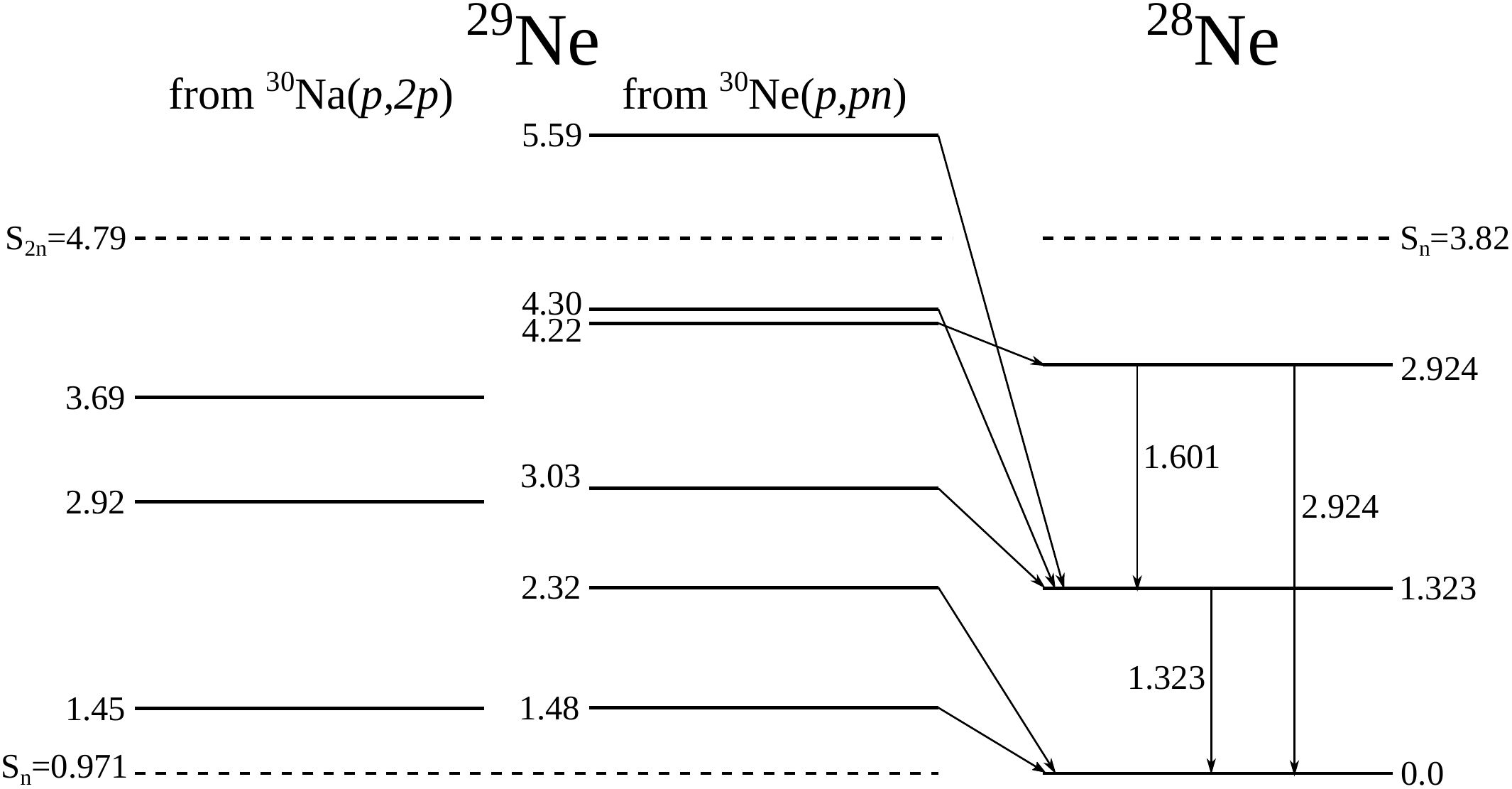}
\caption{Partial level scheme of $^{29}$Ne and $^{28}$Ne. Note that lack of statistics prevented a detailed analysis of $\gamma$-rays in coincidence with the unbound states observed in $^{30}\mathrm{Na}(p,2p)$, see text, so only transitions from states observed in $^{30}\mathrm{Ne}(p,pn)$ are shown. The bound states of $^{29}$Ne are omitted. For $^{28}$Ne, only transitions relevant to the analysis presented here are indicated \cite{Wang2021}.}
\label{fig:level_scheme_simplified}
\end{figure}

The $\gamma$-ray energy $E_{\gamma}$ spectrum measured with DALI2 in coincidence with a $^{28}$Ne fragment and a neutron from $^{30}\mathrm{Ne}(p,pn)$ is shown in Fig. \ref{fig:gamma}. To determine the relative strength of the transitions, the spectrum was fitted using a combination of simulated line shapes and an exponential background. In total eight transitions were included in the fit. A detailed analysis of the energies of these transitions in $^{28}$Ne will be presented in a separate publication \cite{Wang2021}. In Fig. \ref{fig:level_scheme_simplified} only the transitions important to the analysis presented here are shown.\\
The upper panel of Fig. \ref{fig:gamma} shows the above-mentioned fit for the total $E_{\gamma}$ spectrum. Two transitions with $E_{\gamma} = 1323$ keV and $E_{\gamma} = 1601$ keV are found to dominate. The 1323 keV transition is the transition from the $^{28}$Ne first $2^+$-state to the ground state, its energy slightly varying from the value adopted in Ref. \cite{ShamsuzzohaBasunia2013}. The second, newly observed transition at 1601 keV has been placed above this state and is mainly feeding into the first excited state \cite{Wang2021}. The decay into the ground state is not negligible, however, as demonstrated by the occurrence of the 2924 keV line shape in Fig. \ref{fig:gamma}.\\
In order to determine which resonances are measured in coincidence with these $\gamma$-ray transitions, the fit outlined above was repeated for $E_{\gamma}$ spectra gated on 50~keV-wide $E_{rel}$ bins. The integral of each simulated line shape was taken and compared to the total number of events in each $E_{rel}$ bin. The resulting ratio is shown in the lower panel of Fig. \ref{fig:gamma} for the three transitions mentioned above. The fraction of events in coincidence with the 1323 keV transition drops around $E_{rel}=0.5$ MeV and $E_{rel}=1.3$ MeV, indicating that these two resonances decay directly to the ground state of $^{28}$Ne. Only the resonance $E_{rel}=0.3$ MeV is observed in coincidence with the lines at 1601 keV and 2924 keV. Based on this, the level scheme from Ref. \cite{Liu2017} has been extended with states above the neutron threshold, as presented in Fig. \ref{fig:level_scheme}.\\
Due to low $\gamma$-ray statistics and the broad overlapping nature of the resonances observed in   $^{30}\mathrm{Na}(p,2p)$ no detailed analysis of the $\gamma$-ray coincidences was done in this case. Comparing the $E_{rel}$ spectra with and without a coincident $\gamma$-ray of 1.3~MeV offers no sign of a particular resonance being more prominent when in coincidence with a $\gamma$-ray. Therefore, the resonances are tentatively assigned the excitation energies $E^*$ of 1.45(4)~MeV, 2.92(18)~MeV and 3.69(12)~MeV.\\
Both in $^{30}\mathrm{Ne}(p,pn)$ and $^{30}\mathrm{Na}(p,2p)$ resonances with $E^*\approx 1.5$~MeV and $E^*\approx 3$~MeV are observed. In case of the resonances at 3 MeV, their different widths ($\Delta\Gamma_r\approx2\sigma$) make it unlikely that they are the same resonance. The resonance at 1.5~MeV is discussed in more detail in the next section. 
\subsection{Momentum Distributions} \label{sec:momentum}
\begin{figure}[t!]
\includegraphics[width=0.49\textwidth]{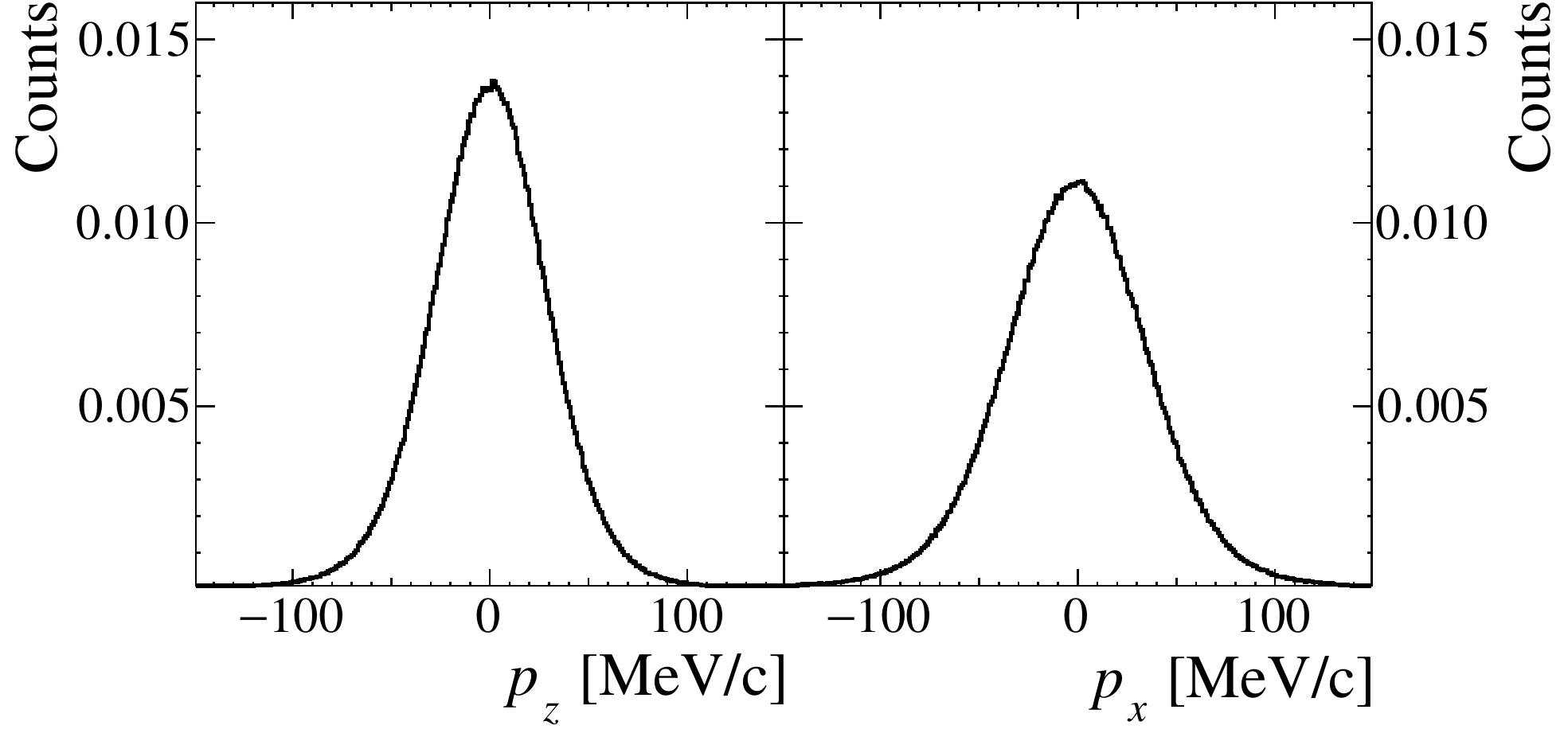}
\caption{Longitudinal (left) and transverse (right) momentum distributions of the unreacted $^{30}$Ne beam. These distributions were used to obtain the detector response for the momentum analysis.}
\label{fig:momentum_res}
\end{figure}
\begin{figure}[t!]
\includegraphics[width=0.49\textwidth]{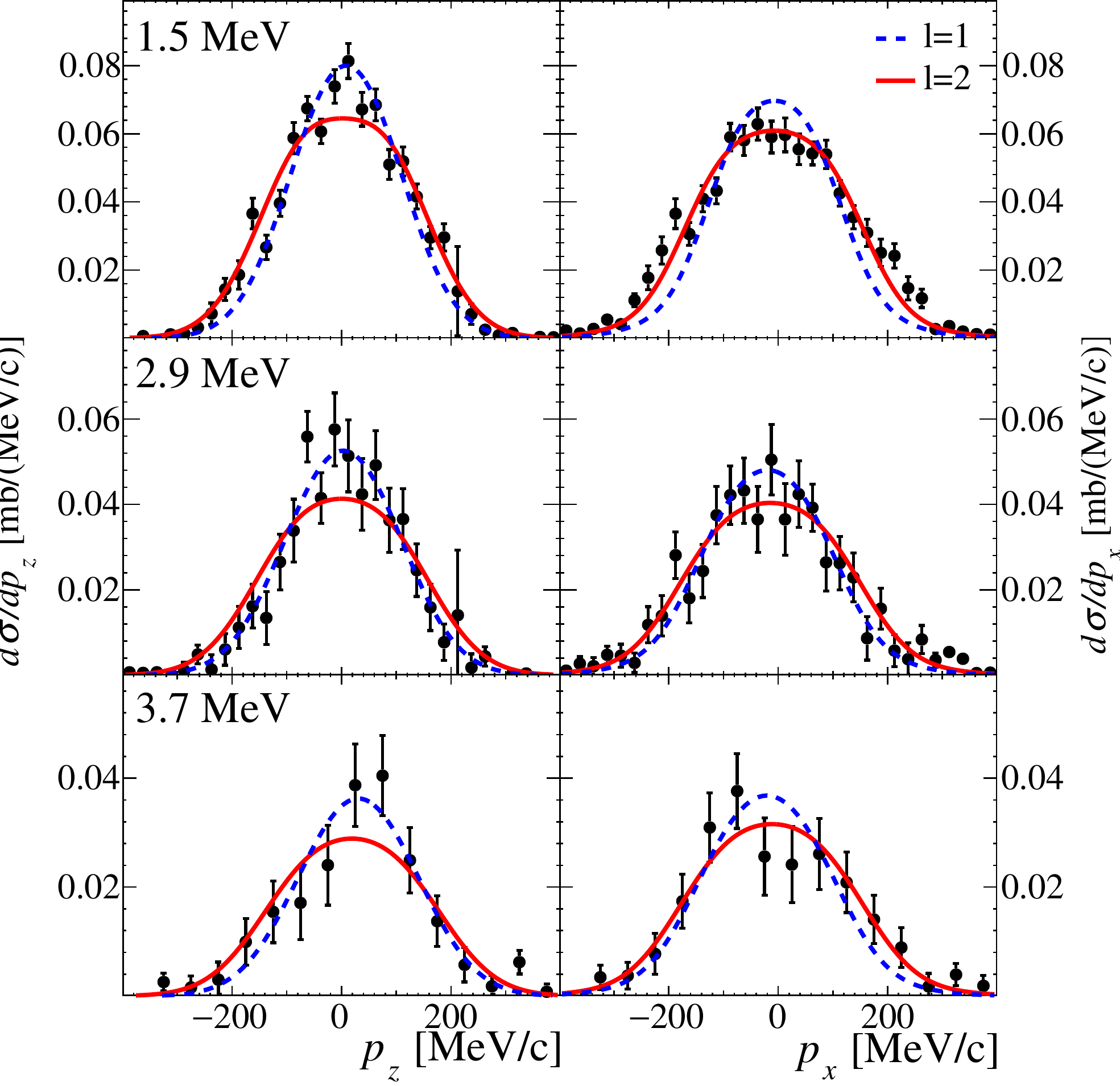}
\caption{Longitudinal (left) and transverse (right) momentum distributions of $^{28}$Ne from $^{30}$Na($p,2p$) without required $\gamma$-ray coincidence. Top row: $E^{*}=1.5$~MeV. Middle row: $E^{*}=2.9$~MeV. Bottom row: $E^{*}=3.7$~MeV.  All distributions have been fitted by predicted shapes obtained with the eikonal model using $\ell$=1 (dashed blue line) and $\ell$=2 (red line) folded with the experimental detector response.}
\label{fig:momentum_p2p}
\end{figure}
\begin{figure}[t!]
\includegraphics[width=0.49\textwidth]{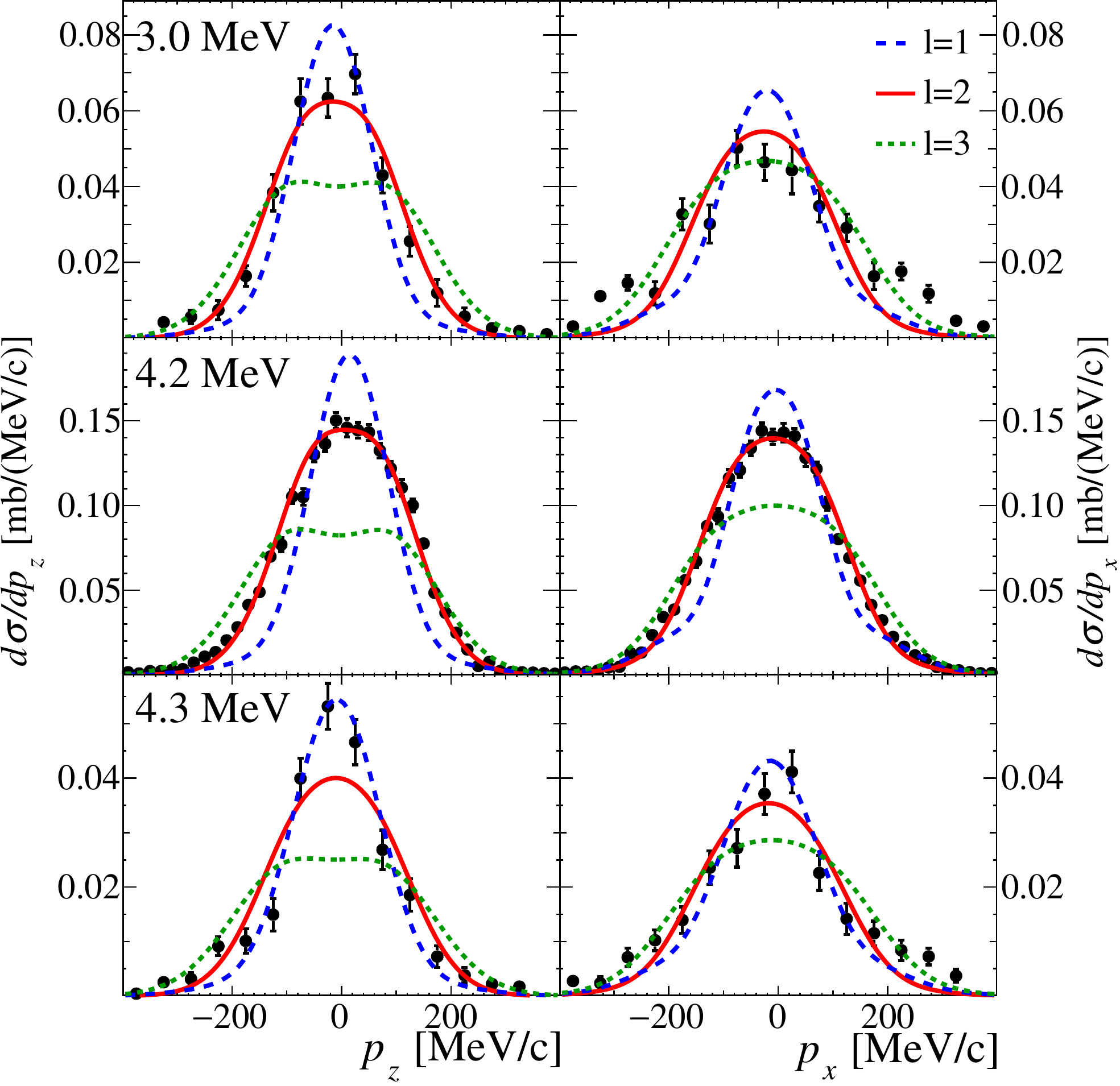}
\caption{Longitudinal (left) and transverse (right) momentum distributions of $^{28}$Ne from $^{30}$Ne($p,pn$) gated on the $E_{\gamma}=1323$~keV transition. Top row: $E^{*}=3.0$~MeV. Middle row: $E^{*}=4.2$~MeV. Bottom row: $E^{*}=4.3$~MeV. All distributions have been fitted with predicted shapes obtained with the eikonal model using  $\ell$=1 (dashed blue line), $\ell$=2 (red solid line) and $\ell$=3 (dotted green line) and folded with the experimental detector response.}
\label{fig:momentum1}
\end{figure}
\begin{figure}[t!]
\includegraphics[width=0.49\textwidth]{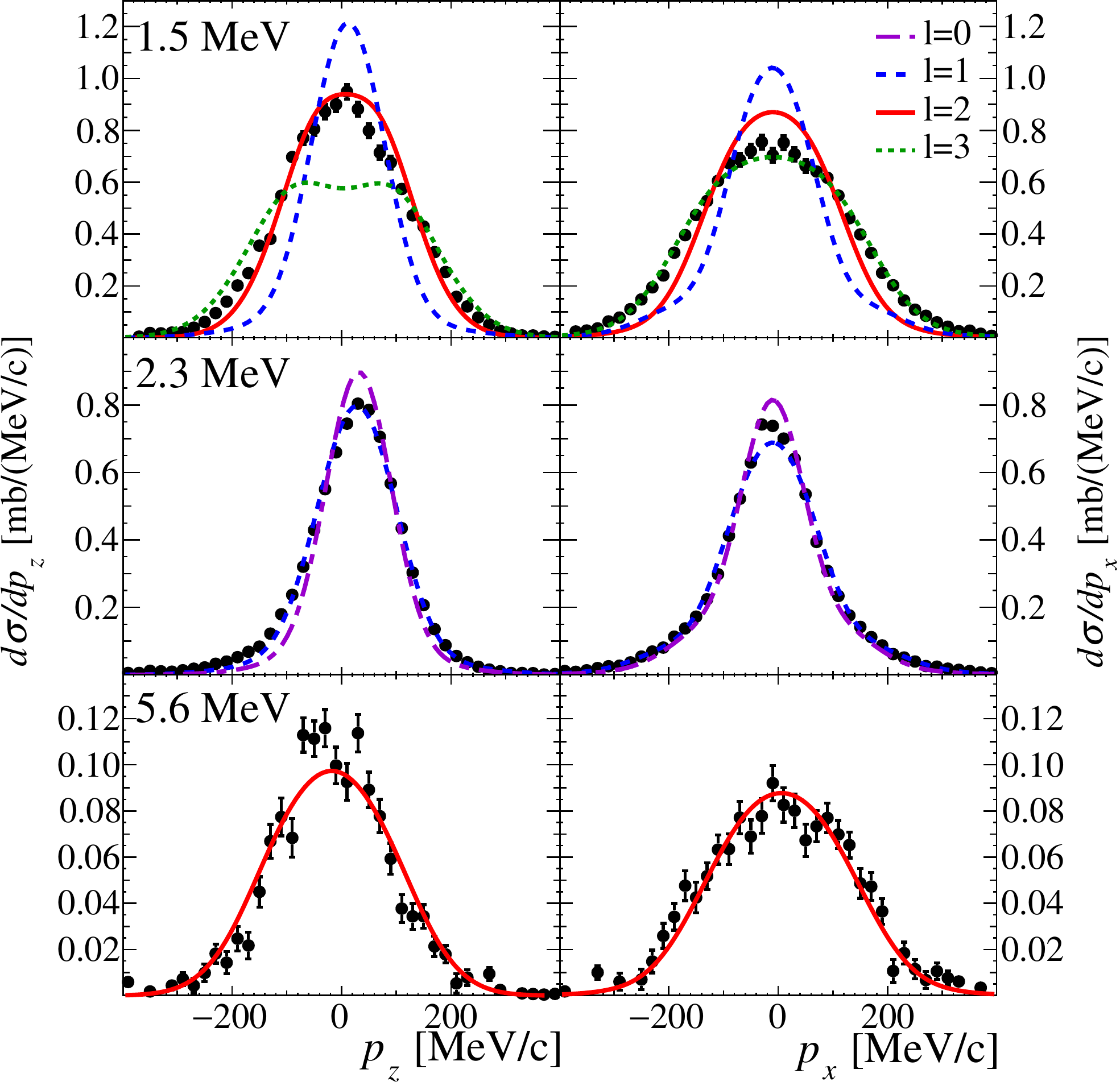}
\caption{Longitudinal (left) and transverse (right) momentum distributions of $^{28}$Ne from $^{30}$Ne($p,pn$) without requiring $\gamma$-ray coincidence compared to predicted shapes from the eikonal model folded with the experimental detector response. Top row: $E^{*}=1.5$~MeV, compared to calculations with $\ell$=1 (blue dashed line), $\ell$=2 (red solid line) and $\ell$=3 (green dotted line). Middle row: $E^{*}=2.3$~MeV, compared to calculations $\ell$=0 (purple dash-dotted line) and $\ell$=1 (blue dashed line). Bottom row: $E^{*}=5.6$~MeV, compared to calculations for $\ell$=2.}
\label{fig:momentum2}
\end{figure}
\begin{figure}[t!]
\includegraphics[width=0.49\textwidth]{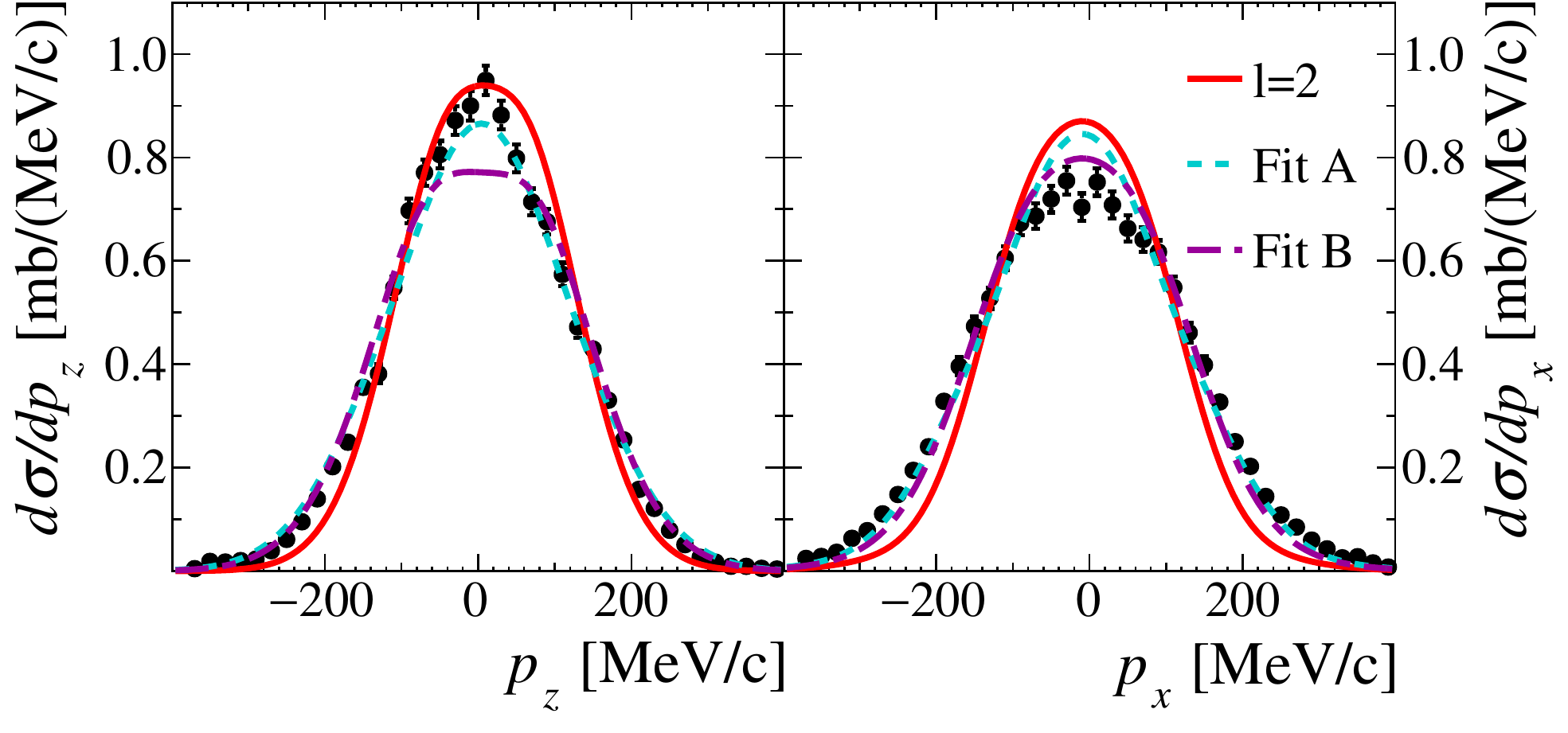}
\caption{Longitudinal (left) and transverse (right) momentum distributions of the lowest-lying observed resonance. The distributions are fitted by linear combinations of $\ell$=1 and $\ell$=3 (Fit A, cyan dashed line), as well as $\ell$=2 and $\ell$=3 (Fit B, purple dash-dotted line) under the assumption that the data reflect the presence of two overlapping resonances. The pure $\ell$=2 distribution is shown for comparison (red line).}
\label{fig:momentum_R1}
\end{figure}
The longitudinal ($p_z$) and transverse ($p_x$) momentum distributions of the $^{28}$Ne fragments were reconstructed from their measured $B\rho$ as well as the angle calculated from the position measurement in FDC1 and the reaction vertex.  The detector response is obtained by measuring the momentum distributions for the unreacted $^{30}$Ne beam. These distributions are shown in Fig. \ref{fig:momentum_res}. The obtained resolution is $\sigma$=28~MeV/c for $p_z$ and $\sigma$=35~MeV/c for $p_x$.\\
For the reacted beam, the momentum distributions were obtained by fitting the $E_{rel}$ spectrum for each $p_x$ or $p_z$ bin and obtaining the integral for each resonance. This allows to separate the momentum distributions of the overlapping resonances and to eliminate the influence of the non-resonant background. We note that the momentum distribution is often analyzed for the system of fragment+$1n$ (in this case $^{29}$Ne) instead of the fragment ($^{28}$Ne) \cite{Revel2020}. However, we find that the difference is small and can be well described using the $^{28}$Ne momentum and incorporating the (compared to the experimental resolution) small additional broadening due to the decay energy of the neutron.\\
The reconstructed longitudinal ($p_z$) and transverse ($p_x$) momentum distributions of the $^{28}$Ne fragments from $^{30}$Na($p,2p$) are shown in Fig. \ref{fig:momentum_p2p}. The distributions are compared to predicted shapes from eikonal model calculations. The direct-reaction model used here for the proton target, analogous to that used extensively for reactions performed on light target nuclei \cite{Hansen2003}, was used previously for the $^{29}\mathrm{Ne}(p,2p)$ and $^{29}\mathrm{F}(p,pn)$ reactions in Ref. \cite{Revel2020}. Details of the model calculations are outlined in Ref. \cite{Revel2020supp} and further formal details as well as the likely parameter sensitivities are presented in Ref. \cite{Tostevin2021}. The calculated curves have been folded with the experimental detector response and the additional broadening due to the decay of the $^{29}$Ne unbound states \cite{Tostevin2001} while their height was fitted to the experimental data.\\
For all three cases, the distributions are compared to the shapes of the momentum distributions calculated with $\ell$=1 and $\ell$=2. The resonance at $E^{*}$=1.5~MeV is very well described by $\ell$=2 for both $p_z$ and $p_x$, as expected for the direct population of positive parity states in $sd$-shell proton removal. For the other two resonances, it is difficult to assign an $\ell$-value unambiguously. This is due to limited statistics for these cases and due to the fact that, since they are well-bound cases, both calculated distributions are relatively broad. However, as the gap between the valence $d_{5/2}$ and filled $p_{1/2}$ orbitals is expected to be large, these resonances are more likely to result from $\ell$=2 proton removal than from fragments of $p$-shell strength at these excitation energies.\\
The momentum distributions of the $^{28}$Ne fragments from $^{30}$Ne($p,pn$) are shown in Figs. \ref{fig:momentum1} and \ref{fig:momentum2}. For the three resonances at $E^{*}$=3.0~MeV, 4.2~MeV and 4.3~MeV, the momentum distributions in coincidence with the observed 1.3~MeV $\gamma$-ray transition are shown in Fig. \ref{fig:momentum1}. Due to statistics and the overlapping nature of the resonances, the angular momentum cannot be assigned completely unambiguously for the resonance at $E^*=3.0$~MeV. However, the best agreement with the data is found for $\ell$=2. The resonance at $E^*=4.2$~MeV is very well described by calculations with $\ell$=2. For the resonance at $E^*=4.3$~MeV, a clear assignment is again difficult, as both $\ell=1$ and $\ell$=2 describe the data well.\\
The momentum distributions of the remaining three observed resonances are shown in Fig. \ref{fig:momentum2}. In these cases, the distributions were not gated on coincident $\gamma$-ray transitions. In the case of the resonances at $E^{*}=1.5$~MeV and 2.3~MeV, this was done because, as discussed in the previous section, these resonances are not in coincidence with any $\gamma$-ray. For the resonance at 5.6~MeV, the $\gamma$-ray gated distributions have too low statistics. The distributions of the 5.6~MeV state can be described very well by calculations with $\ell$=2. The narrow distribution for $E^*=2.3$~MeV is best described by calculations with $\ell$=1. Calculations with $\ell$=0 are also shown for comparison, but found to be slightly too narrow, especially for $p_z$.\\
In the case of the 1.5~MeV state, none of the individual $\ell$ calculations describes both the longitudinal and the transverse distribution. In particular, the calculated curves with $\ell$=2 are found to be too narrow, despite the fact that a positive-parity resonance at about the same excitation energy, which can be described by $\ell$=2, was found in $^{30}\mathrm{Na}(p,2p)$.\\
The distributions were therefore fitted by linear combinations of $\ell$=1 and $\ell$=3 (Fit A, cyan dashed line), as well as $\ell$=2 and $\ell$=3 (Fit B, purple dash-dotted line) under the assumption that the data reflect the presence of two overlapping resonances. These fits are shown in Fig. \ref{fig:momentum_R1}. The longitudinal and transverse momentum distributions were fitted simultaneously, with the same ratio between the different $\ell$-components for both fits. The relative contribution of the $\ell$=3-component was found to be 71\% for Fit A and 54\% for Fit B. While Fit A describes the longitudinal momentum distribution better, Fit B gives the better description of the transverse momentum. Overall the best agreement is found with Fit A ($\chi^2/N=5.4$ compared to $\chi^2/N=6.6$ for Fit B). A fit of a pure $\ell$=2-calculation is also shown for comparison (green dotted line). The resulting curve is too narrow to describe the data for both $p_z$ and $p_x$ ($\chi^2/N=24.0$). As $\ell$=0 or $\ell$=1 would lead to even narrower distributions, this is a clear indication that the data cannot be explained without a strong $\ell$=3 contribution.\\

\section{Comparison with Shell Model Calculations}
\begin{figure}[t]
\includegraphics[width=0.49\textwidth]{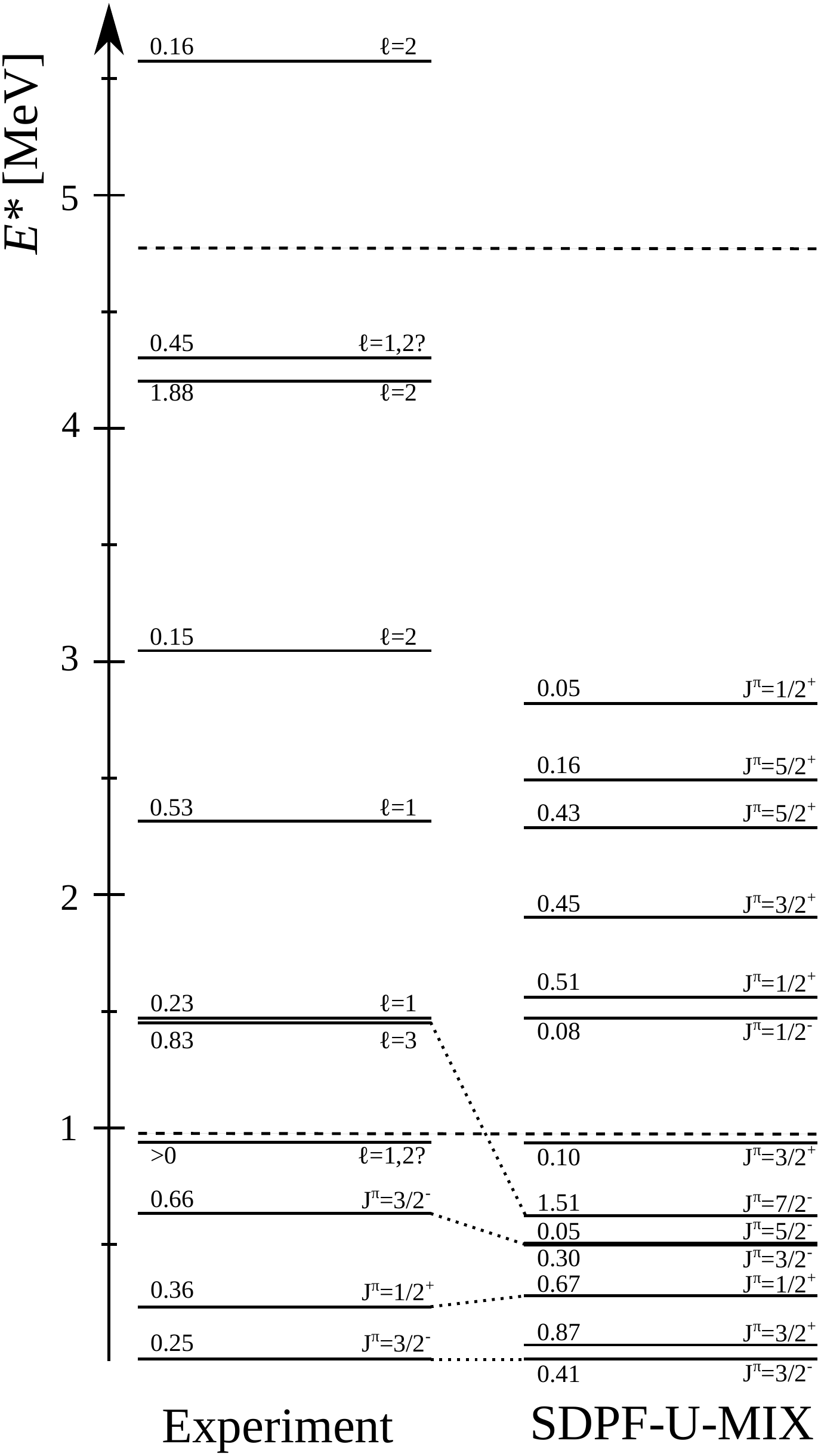}
\caption{Partial level scheme of $^{29}$Ne. The scheme from Ref. \cite{Liu2017} has been extended to states above the neutron threshold observed in this work. The assigned angular momentum and extracted spectroscopic factor is indicated for each state. For clarity only unbound states observed in $^{30}\mathrm{Ne}(p,pn)$ are included in the figure. The energy levels from shell model calculations with $C^2S\geq0.05$ for $1n$-removal from the ground state of $^{30}$Ne using the {\sc sdpf-u-mix} interaction are shown for comparison. The one-neutron and two-neutron separation energies are indicated by the dashed lines.}
\label{fig:level_scheme}
\end{figure}
\begin{figure}[t]
\includegraphics[width=0.49\textwidth]{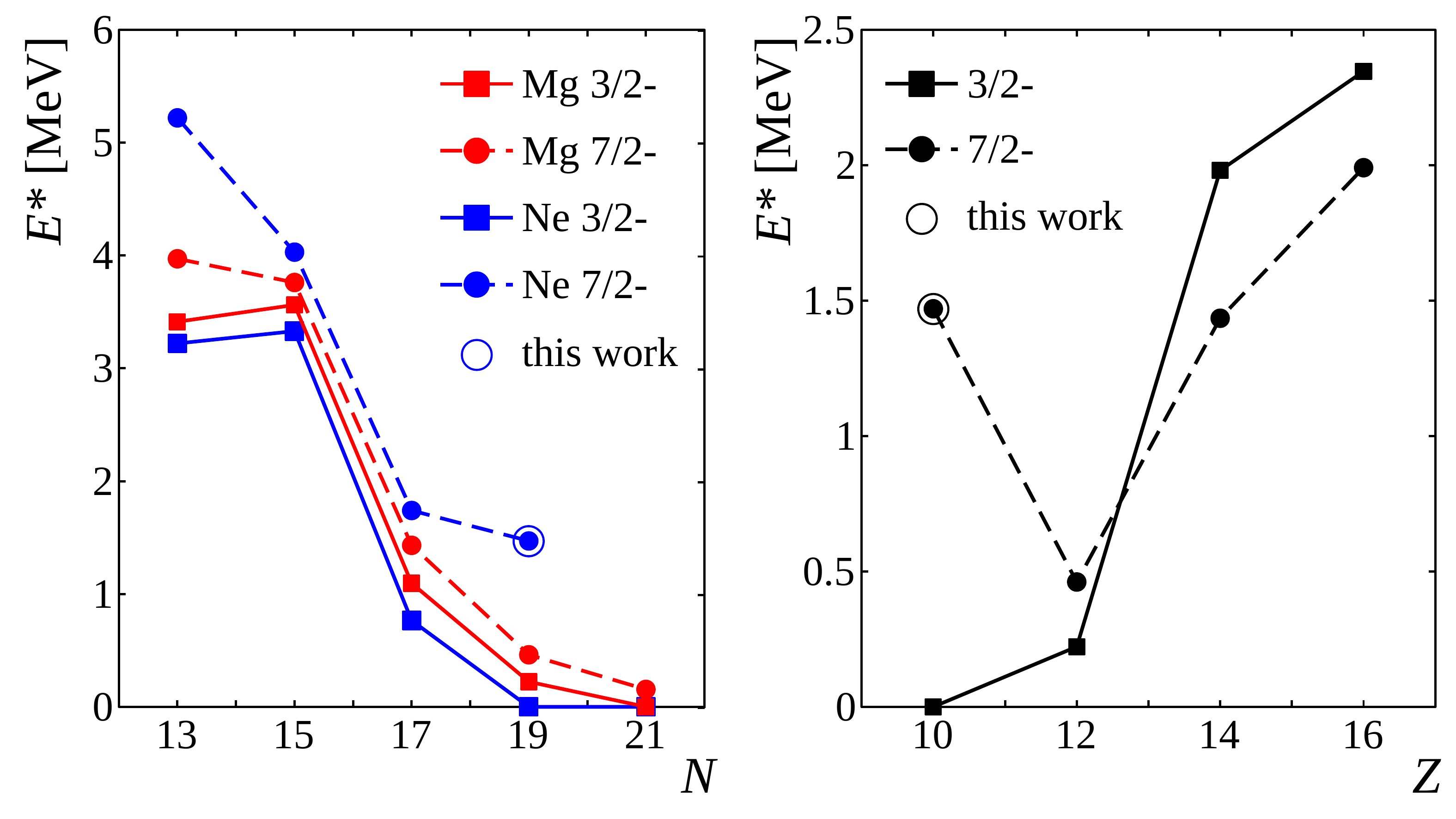}
\caption{Left: Excitation energy for the lowest-lying $3/2^-$ and $7/2^-$ states in neutron-rich odd Mg and Ne isotopes. The values for the Mg isotopes have been taken from Refs. \cite{Firestone2009,ShamsuzzohaBasunia2011,Matta2019,Terry2008,Bazin2021}, the values for Ne from Refs. \cite{ShamsuzzohaBasunia2021,Firestone2009,Catford2010,Brown2012,Liu2017,Nakamura2014}. Right: Excitation energies for the lowest-lying $3/2^-$ and $7/2^-$ states in $N$=19 isotones. Values taken from Refs. \cite{Liu2017,Terry2008,Jongile2020,Chen2011a}.}
\label{fig:exe_vs_n}
\end{figure}
\begin{table*}[t]
\caption{Observed states of $^{29}$Ne from  $^{30}$Ne($p,pn$) and $^{30}$Na($p,2p$). The spectroscopic factors $C^2S=\sigma_{exp}/\sigma_{sp}$ are calculated using eikonal single-particle cross sections $\sigma_{sp}$ and the $\ell$-assignments from the momentum distribution analysis. For the resonance at $E^*$=1.48~MeV, the spectroscopic factors are given for both fits. For states with assignment $\ell$=2, the $C^2S$ is calculated using $\sigma_{sp}$ for both $d_{3/2}$ and $d_{5/2}$. The error shown for $C^2S$ is based on experimental uncertainties only.}
\label{tab:c2s}
\begin{tabular}{p{.2\textwidth} p{.1\textwidth} p{.1\textwidth} p{.1\textwidth} D{.}{.}{2.8} p{.1\textwidth} p{.1\textwidth} p{.1\textwidth}} 
 & $E^*$ [MeV] & $\Gamma_r$ [MeV] & $\ell$ & \multicolumn{1}{p{.1\textwidth}}{$\sigma$ [mb]} & $\ell_j$ & $\sigma_{sp}$ [mb]& $C^2S$ \\ 
 \hline
 $^{30}\mathrm{Ne}(p,pn)^{29}\mathrm{Ne}^*$& 1.48(4) & 0.17(8) & $1+3$ & 12.9(5) &  &  & \\  
 &         &          &       & 3.7(2) & $p_{3/2}$ & 15.29 & 0.23(1) \\  
 &         &          &       & 9.2(4) & $f_{7/2}$ & 11.12 & 0.83(4) \\  
 & 1.48(4) & 0.17(8) & $2+3$ & 12.9(5) &  &  & \\  
 &         &          &       &  5.9(3) & $d_{5/2}$ & 12.11 & 0.49(2) \\  
 &         &          &       &  7.0(3) & $f_{7/2}$ & 11.12 & 0.63(3) \\  
 & 2.32(2) & 0.30(4)  & 1     & 7.8(4)  &  &  &  \\  
 &         &          &       &  & $p_{3/2}$ & 14.64 & 0.53(3) \\  
 & 3.03(7) & 0.11(5)  & 2     & 1.5(6)  &  &  &  \\  
 &         &          &       &  & $d_{3/2}$ & 10.30 & 0.15(6) \\  
 &         &          &       &  & $d_{5/2}$ & 11.30 & 0.13(5) \\  
 & 4.22(2) & 0.05(1)  & 2     & 18.5(5) &  &  &  \\  
 &         &          &       &  & $d_{3/2}$ &  9.85 & 1.88(4) \\  
 &         &          &       &  & $d_{5/2}$ & 10.80 & 1.71(4) \\  
 & 4.30(1) & 0.33(12) & (1,2) & 4.3(8)  &  &  &  \\  
 &         &          &       &   & $p_{3/2}$ & 13.50 & 0.32(5)\\  
 &         &          &       &   & $d_{3/2}$ &  9.85 & 0.45(8)\\  
 &         &          &       &   & $d_{5/2}$ & 10.80 & 0.40(7) \\  
 & 5.60(8) & 0.7(3)   & 2     & 1.5(4)  &  &  &  \\
 &         &          &       &   & $d_{3/2}$ &  9.42 & 0.16(4) \\
 &         &          &       &   & $d_{5/2}$ & 10.32 & 0.15(4) \\
 \hline
 $^{30}\mathrm{Na}(p,2p)^{29}\mathrm{Ne}^*$& 1.45(4){\textsuperscript a}  & 0.05(3)  & 2     & 0.9(2)  & $d_{5/2}$ & 4.53 & 0.20(3) \\
 & 2.92(18){\textsuperscript a}  & 0.8(4)  & 2     & 0.6(2)  & $d_{5/2}$ & 4.42 & 0.13(4) \\
 & 3.69(12){\textsuperscript a}  & 0.6(3)  & 2     & 0.2(1)  & $d_{5/2}$ & 4.36 & 0.06(3) \\
 \hline
\multicolumn{8}{l}{{\textsuperscript a}Possible coincident $\gamma$-rays were not analyzed.}
\end{tabular}
\end{table*}

The level scheme of $^{29}$Ne obtained from $^{30}$Ne($p,pn$) is compared to shell model calculations for the $1n$-removal from the ground state of $^{30}$Ne using the {\sc sdpf-u-mix} interaction \cite{Caurier2014} in Fig. \ref{fig:level_scheme}. The spectroscopic factors $C^2S$ are also indicated. The experimental spectroscopic factors are determined as $C^2S=\sigma_{exp}/\sigma_{sp}$ using eikonal single-particle cross sections $\sigma_{sp}$ based on the $\ell$-assignments from the momentum distribution analysis \cite{Tostevin2021}. The theoretical levels and spectroscopic factors have been calculated using the code ANTOINE \cite{Caurier2005}. Only levels with $C^2S\geq0.05$ are shown.\\
Theoretically, the $2^+$ ground state of $^{30}$Na and the $0^+$ ground state of $^{30}$Ne are dominated by $2p$-$2h$ and $2p$-$2h$/$4p$-$4h$ neutron configurations, respectively \cite{Poves}, with the expectation of significant occupancy of the $1f_{7/2}$ orbital - 1.7 in the case of the {\sc sdpf-u-mix} shell-model calculation for $^{30}$Ne. It is the location of such $7/2^-$ removal strength in the $^{29}$Ne spectrum that is of interest here.\\
The observed states of $^{29}$Ne  and their spectroscopic factors $C^2S$ are also listed in Table \ref{tab:c2s}. For the resonance at $E^*$=1.48~MeV, the spectroscopic factors for both fits are given. For states with assignment $\ell$=2, the spectroscopic factors are calculated using $\sigma_{sp}$ for both $d_{3/2}$ and $d_{5/2}$ as the data do not allow to differentiate between these two cases. The error shown for $C^2S$ is based on experimental uncertainties only.\\
In the present data, a single $\ell=3$ state is found at the excitation energy of $E^*=1.48$~MeV. Available effective interactions predict the lowest-lying $7/2^-$-state to be bound, at 610~keV, in the case of {\sc sdpf-u-mix}. However, despite a large predicted cross section for this state, it was missing in the observed $\gamma$-ray spectrum following $^{30}\mathrm{Ne}(p,pn)$ to bound states \cite{Liu2017}. The state at  $E^*=1.48$~MeV is therefore assigned to be the first state dominated by an $f_{7/2}$-hole. The large observed spectroscopic strength for this state is also comparable to the predictions by the shell model calculations for the lowest-lying $7/2^-$-state, supporting this assignment.\\
The difference between predicted and measured excitation energy is noteworthy, given that {\sc sdpf-u-mix} was found to provide otherwise rather good agreement with experimental data to bound $^{29}$Ne final states in Ref. \cite{Liu2017}. It is interesting if this discrepancy is inherent to the continuum nature of the state. If so, further theoretical work including the continuum effects such as Gamow shell model would be useful to understand the properties of the intruder states in the island-of-inversion nuclei.\\ 
No negative-parity state of sufficient strength is predicted in the vicinity of the $\ell$=1-resonance at 2.3~MeV. The calculations shown do not extend to the excitation energy of $\approx 4.2$~MeV where strong $\ell$=2 resonances are observed. However, they predict strong positive parity states already at about $E^*=2$~MeV.\\
To put this result in perspective in terms of the border of the island of inversion, the left panel of Fig. \ref{fig:exe_vs_n} shows the excitation energies of the lowest-lying $3/2^-$ and $7/2^-$ states for neutron-rich odd Mg and Ne isotopes as a function of the neutron number. The newly measured excitation energy in $^{29}$Ne breaks with the trend observed for the $3/2^-$-states as well as for the $7/2^-$-states in Mg, where the excitation energies are lowered significantly for $N$=19 compared to $N$=17. In the right panel of Fig. \ref{fig:exe_vs_n}, the  excitation energies of the lowest-lying $3/2^-$ and $7/2^-$ states are shown for the $N$=19 isotones. In both cases, a significant decrease is seen when going from $Z$=14 to $Z$=12. However, while the energy for the $3/2^-$-state decreases even further when going to $N$=10, our newly obtained value for $7/2^-$ states shows $E^*$ increasing again.\\

\section{Conclusion}
In summary, unbound states of $^{29}$Ne have been measured and analyzed for the first time using the nucleon removal reactions $^{30}\mathrm{Ne}(p,pn)^{29}\mathrm{Ne}^*$ and $^{30}\mathrm{Na}(p,2p)^{29}\mathrm{Ne}^*$. The combined information from resonance energies, partial cross sections and momentum distributions allowed us to identify and characterize a total of 9 unbound states in $^{29}$Ne. The momentum distributions of the lowest-lying resonance at $E^*=1.48$~MeV cannot be explained without an $\ell$=3-component, providing first evidence of the location of the low-lying $f_{7/2}$ single-particle strength in $^{29}$Ne. This is also supported by the large observed spectroscopic strength for this state which is also predicted for the $f_{7/2}$-state by shell model calculations using the {\sc sdpf-u-mix} interaction. However, the absolute energy of the $7/2^-$-state is found to be higher by 0.9 MeV compared to the model prediction. While the 3/2$^-$ ground state of $^{29}$Ne places this nucleus into the island of inversion at $N$=20, the findings of this work confirm that the lowest-lying $f_{7/2}$ intruder state is unbound. The splitting of the $fp$-shell intruder states increases in contrast to lighter neon isotopes.\\
The properties of the resonances observed in $^{30}\mathrm{Na}(p,2p)$ are consistent with removal from the $d_{5/2}$-orbital, as expected for the removal of a deeply-bound proton from $^{30}$Na.

\section*{Acknowledgements}
We thank Alfredo Poves and Fr\'ed\'eric Nowacki for providing results from large-scale shell model calculations. Support was provided by the Swedish Research Council under grant numbers 2011-5324 and 2017-03839. This work was supported partially by JSPS KAKENHI Grant Nos. JP16H02179, and JP18H05404. JAT acknowledges visiting researcher support at Tokyo Institute of Technology from MEXT KAKANHI Grant No. JP18H05400 and the support of the Science and Technology Facilities Council (U.K.) Grant No. ST/L005743/1. The LPC-Caen group wished to acknowledge support from the Franco-Japanese LIA-International Associated Laboratory for Nuclear Structure Problems as well as the French ANR-14-CE33-0022-02 EXPAND. This project was partly funded by the Deutsche Forschungsgemeinschaft (DFG, German Research Foundation) - Project-ID 279384907 - SFB 1245 and the GSI-TU Darmstadt cooperation agreement.  This work was in part supported by the Institute for Basic Science (IBS-R031-D1) in Republic of Korea. Additional support from the European Regional Development Fund contract No. GINOP-2.3.3-15-2016-00034 and the National Research, Development and Innovation Fund of Hungary via Project No. K128947. I.G. has been supported by HIC for FAIR and Croatian Science Foundation under projects no. 1257 and 7194. Y.T. acknowledges the support by JSPS Grant-Aid for Scientific Research Grants No. JP21H01114. This material is based upon work supported by the U.S. Department of Energy, Office of Science, Office of Nuclear Physics, under Contract No. DE-AC02-06CH11357 (ANL).

\bibliographystyle{apsrev4-2}
\bibliography{references}
\end{document}